%% file: main.tex
\documentclass[sigconf, screen]{acmart}

\usepackage{ifthen}
\usepackage{xcolor}
\usepackage{amsmath}
\usepackage{enumitem}

\usepackage{amsthm}

\theoremstyle{definition}
\newtheorem{definition}{Definition}[section]

\newtheorem{theorem}{Theorem}[section]

\newtheorem{lemma}{Lemma}[section]

\usepackage{times} 
\usepackage{amsfonts}
\usepackage{mathrsfs}
\usepackage{dsfont}
\usepackage{amsbsy}
\usepackage{stmaryrd}
\usepackage{url}
\usepackage{xpatch}

\usepackage{graphicx}

\usepackage{pict2e}

\DeclareRobustCommand{\diamonddiamond}{%
  \begingroup
  \setlength{\unitlength}{\fontcharht\font`T}%
  \begin{picture}(1,1)
  \polygon(.5,0)(1,.5)(.5,1)(0,.5)
  \polygon(.5,.25)(.75,.5)(.5,.75)(.25,.5)
  \end{picture}%
  \endgroup
}

\usepackage{color}

\definecolor{light-gray}{gray}{0.85}
\colorlet{shadecolor}{light-gray}
\usepackage{framed}

\usepackage{algorithm}
\usepackage{algpseudocode}
\usepackage{circledsteps}

\usepackage{subcaption}
\usepackage{graphicx}
\usepackage{multirow}

\usepackage{hyperref}

\usepackage{makecell}

\algnewcommand\algorithmicswitch{\textbf{switch}}
\algnewcommand\algorithmiccase{\textbf{case}}
\algnewcommand\algorithmicassert{\texttt{assert}}
\algnewcommand\Assert[1]{\State \algorithmicassert(#1)}%
\algdef{SE}[SWITCH]{Switch}{EndSwitch}[1]{\algorithmicswitch\ #1\ \algorithmicdo}{\algorithmicend\ \algorithmicswitch}%
\algdef{SE}[CASE]{Case}{EndCase}[1]{\algorithmiccase\ #1}{\algorithmicend\ \algorithmiccase}%
\algtext*{EndSwitch}%
\algtext*{EndCase}%

\newboolean{showcomments}
\setboolean{showcomments}{false} 

\ifthenelse{\boolean{showcomments}}
{\newcommand{\nb}[3]{
  \fcolorbox{black}{#3}{\bfseries\sffamily\scriptsize#1}
  {\sf\small\textit{#2}$\blacktriangleleft$}
 }
 
}
{\newcommand{\nb}[3]{}
 
}

\newcommand{\ev}[1]{\mathrm{F}_{#1}}
\newcommand{\glob}[1]{\mathrm{G}_{#1}}
\newcommand{\until}[1]{\mathrm{U}_{#1}}

\newcommand{\somewhere}[2]{\diamonddiamond_{#1}^{#2}}
\newcommand{\everywhere}[2]{\boxbox_{#1}^{#2}}

\newcommand{\reach}[2]{\mathcal{R}_{#1}^{#2}}
\newcommand{\escape}[2]{\mathcal{E}_{#1}^{#2}}

\AtBeginDocument{%
  \providecommand\BibTeX{{%
    \normalfont B\kern-0.5em{\scshape i\kern-0.25em b}\kern-0.8em\TeX}}}

\setcopyright{acmcopyright}
\copyrightyear{2021}
\acmYear{2021}
\acmDOI{---------------}

\acmConference[MEMOCODE '21]{MEMOCODE '21: 19th ACM-IEEE
International Conference on Formal Methods and Models for System Design}{November 20--22, 2021}{Beijing, China}
\acmBooktitle{MEMOCODE '21: ACM-IEEE
International Conference on Formal Methods and Models for System Design,
  November 20--22, 2021, Beijing, China}
\acmPrice{15.00}
\acmISBN{-------------}

\begin{document}

\title[Online Monitoring of Spatio-Temporal Properties for Imprecise Signals]{Online Monitoring of Spatio-Temporal Properties for Imprecise Signals}

\author{Ennio Visconti}
\affiliation{%
  \institution{TU Wien}
  \streetaddress{Treitlstraße 3}
  \city{Vienna}
  \country{Austria}}

\author{Ezio Bartocci}
\affiliation{%
  \institution{TU Wien}
  \streetaddress{Treitlstraße 3}
  \city{Vienna}
  \country{Austria}}

\author{Michele Loreti}
\affiliation{%
 \institution{Università di Camerino}
 \city{Camerino}
 \country{Italy}}

\author{Laura Nenzi}
\affiliation{%
  \institution{TU Wien}
  \streetaddress{Treitlstraße 3}
  \city{Vienna}
  \country{Austria}}
\affiliation{%
  \institution{Università di Trieste}
  \city{Trieste}
  \country{Italy}}

\renewcommand{\shortauthors}{Visconti, et al.}

\input{sections/abstract}

\ccsdesc[500]{Theory of computation~Logic and verification}
\ccsdesc[500]{Theory of computation~Modal and temporal logics}
\ccsdesc[300]{Software and its engineering~Abstraction, modeling and modularity}

\keywords{Runtime verification, online monitoring, spatio-temporal logic, imprecise signal, signal temporal logic}


\maketitle

\input{sections/intro}

\input{sections/background}
\input{sections/logic}
\input{sections/monitoring}

\input{sections/eval}
\input{sections/future}

\begin{acks}
The authors would like to acknowledge Davide Prandini for his thesis work (unpublished) where a preliminary work on imprecise signals for STL had been conducted, together with many ideas that have been used for developing the proofs of the theorems presented. This research has been partially supported by the Austrian FWF projects ZK-35 and LogiCS DK W1255-N23; and by Italian MIUR project PRIN 2017FTXR7S IT MATTERS and by Marche Region in implementation of the financial programme POR MARCHE FESR 2014-2020, project "Miracle". 
\end{acks}

\bibliographystyle{ACM-Reference-Format}
\bibliography{biblio/ennio}

\clearpage
\appendix

\input{sections/proofs}

\input{sections/extra_algo}

\end{document}

%% file: sections/abstract.tex
\begin{abstract}
From biological systems to cyber-physical systems, monitoring the behavior of such dynamical systems often requires to reason about complex spatio-temporal properties of physical and/or computational entities that are dynamically interconnected and arranged in a particular spatial configuration. 
Spatio-Temporal Reach and Escape Logic (STREL) is a recent logic-based formal language designed to specify and to reason about spatio-temporal properties. STREL considers each system's entity as a node of a dynamic weighted graph representing their spatial arrangement.  Each node generates a set of mixed-analog signals describing the evolution over time of computational and physical quantities characterising the node's behavior. While there are offline algorithms available for monitoring STREL specifications over logged simulation traces, here we investigate for the first time an online algorithm enabling the runtime verification during the system's execution or simulation.
Our approach extends the original framework by considering imprecise signals and by enhancing the logics' semantics with the possibility to express partial guarantees about the conformance of the system's behavior with its specification. 
Finally, we demonstrate our approach in a real-world environmental monitoring case study.


\end{abstract}

%% file: sections/intro.tex
\section{Introduction}\label{sec:intro}
Complex emergent spatio-temporal patterns such as 
traffic congestion or travelling waves
are central in the understanding of networked dynamical systems
where locally interacting entities are
operating at different time and spatial scales.
We can observe these patterns both in biological systems~\cite{GrosuSCWEB09,BartocciBMNS15,Bartocci2016TNCS} as well as human engineered artefacts such as Collective Adaptive Systems~\cite{LoretiH16} (CAS) and Cyber-Physical Systems~\cite{RatasichKGGSB19} (CPS).
 CAS and CPS consist of a large number of heterogeneous
 (physical and computational in CPS) and spatially distributed entities featuring complex interactions among themselves, with humans and other systems.  Example include biking sharing 
 systems, the internet of things, contact tracing devices preventing the epidemic spread, vehicular networks and smart cities.  Many of these systems are also safety-critical~\cite{RatasichKGGSB19}, meaning that a failure could result in loss of life or in catastrophic consequences for the environment. 
 
The complex interaction with the physical environment
in which these systems are embedded prevents them from being  exhaustively verified at design time. A common alternative is testing~\cite{chapter5}: 
 traces generated during their execution/simulation are stored
 and monitored offline with respect 
 to a formal specification used as an oracle.
 However, testing may provide a limited coverage and does not take into account physical failures that may happen during the execution. Online monitoring is instead a preferable solution when the monitoring verdict requires the immediate action of a policy maker during the system's execution or when it is very computationally expensive~\cite{SelyuninJNRHBNG17} generating and storing the system's execution traces to be monitored offline.   
 \begin{figure*}[!t]
\centering
\includegraphics[width=0.82\paperwidth]{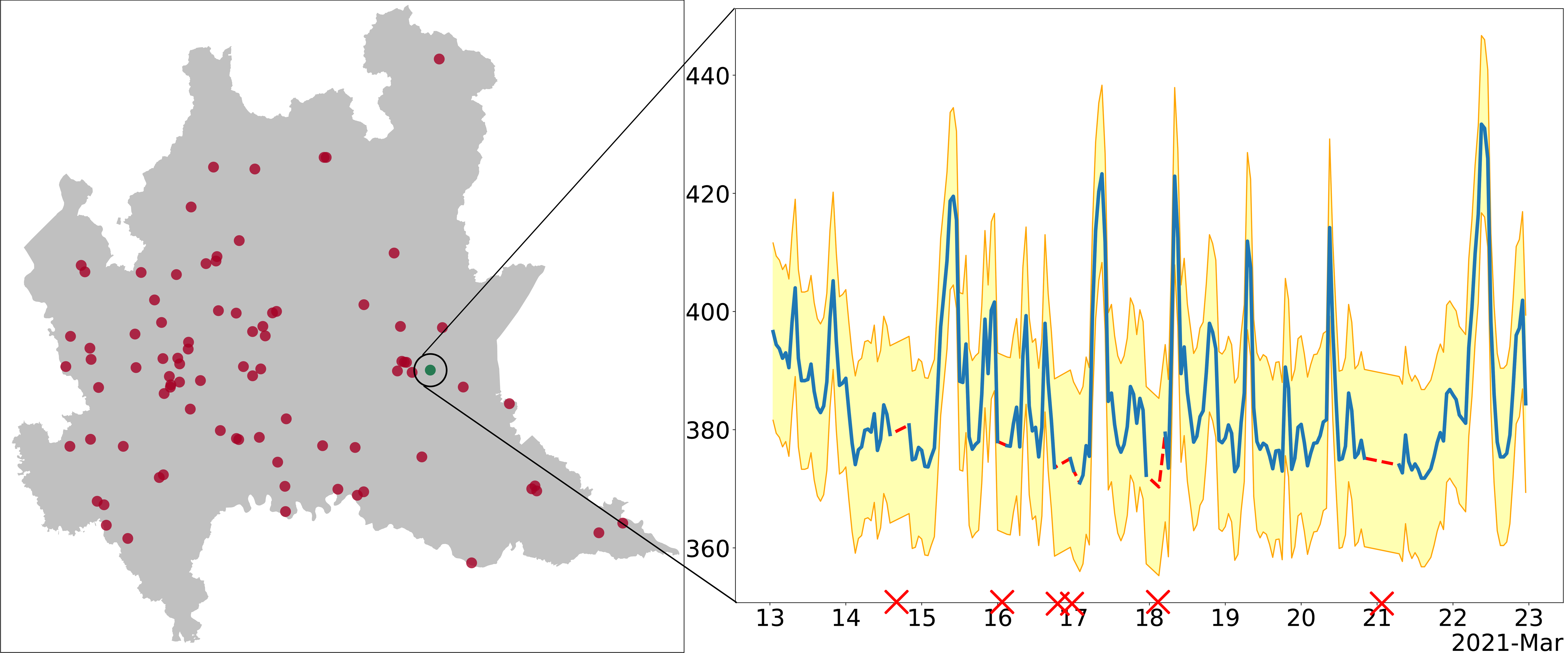}
\caption {
    On the left the NO2 measuring stations of Lombardy; the lighter-colored dot represents the station 6859 - Rezzato. On the right the detailed view of the NO2 measured by the Rezzato station~\cite{arpa-measurements} in terms of $\mu g/m^3$. Red crosses and dotted line represent missing values. Data from the open data initiative of ARPA Lombardia~\cite{arpa-stations}.
  }
\label{fig:no2}
\end{figure*}
 
 
 \noindent \emph{\bf Motivating  example.} As a case study we consider a sensor network for environmental monitoring.
 Air pollution is the primary cause of 
 the loss of biodiversity, of the reduction of agricultural productivity and of many diseases for humans’ lungs and cardiovascular system. Policy makers are constantly monitoring the amount of $NO_2$, an air pollutant that forms from the combustion of fossil fuels, to activate special policies that mitigate its release when levels grow too significantly for the public concern.
For example, in the Italian region of Lombardy, there are over 80 stations distributed throughout the region monitoring the level of $NO_2$ in the air.  Figure~\ref{fig:no2} shows the location of these stations, and the value reported in the town of Rezzato, for the first quarter of 2021.  The measurements happen regularly every hour, but sometimes the sensors fail to communicate the measurements, because of meteorological issues, or temporary faults, and the actual values are provided later on.  Furthermore, the values measured by the  sensors are noisy and they have a certain degree of uncertainty.
Another way to deal with the missing values, could be to check that nearby locations (e.g. within 10 Km) do not register alarming values of particles in the air and this is possible only if we consider spatio-temporal properties.

In this paper \emph{we address
the problem of online monitoring of spatio-temporal properties over systems from which we can observe noisy signals with possible missing data or out-of-order samples.}
 
 \noindent {\bf Spatio-temporal  monitoring.} In the last decade there has been a great effort to develop logic-based specification languages and monitoring frameworks for  spatio-temporal properties. Examples include  SpaTeL~\cite{spatel},  SSTL~\cite{NenziBCLM15},  (SaSTL)~\cite{MaBLS020,sastl2021} and STREL~\cite{Bartocci17memocode}. For more details on the underlying spatial models and the language  expressiveness we refer the reader to~\cite{NenziBBLV20}. 
  In this paper, we consider STREL~\cite{Bartocci17memocode} a spatio-temporal logic operating over a dynamic weighted graph representing the spatial arrangement of spatially distributed entities.  Each node generates a set of mixed-analog signals describing the evolution over time of computational and physical quantities characterising the node's behavior. STREL extends the Signal Temporal Logic (STL)~\cite{MalerN13} with the \emph{reach} and \emph{escape} operators that generalizes the
 \emph{somewhere, everywhere} and \emph{surronded} spatial modalities, simplifying the monitoring that can 
 be computed locally with respect to each node.
 However, the original work on STREL~\cite{Bartocci17memocode,BartocciBLNS20}  provides only an offline monitoring algorithm.  In contrast we present here the first online monitoring algorithm for STREL and in general for spatio-temporal monitoring.  

\noindent {\bf Online Monitoring.} To the best of our knowledge, the only online monitoring techniques~\cite{online-fainekos,Deshmukh2017, JaksicBGKNN15,JaksicBGN18,rtamt,Mamouras2020,MamourasCW21} that are available in the literature can handle temporal specification languages such as STL~\cite{MalerN13} and Metric Temporal Logic~\cite{Koymans90} (MTL).  One of the main challenges for online monitoring is how and when to decide the satisfaction/violation of a formula with temporal operators reasoning about future and not yet observed events. 
In~\cite{online-fainekos} the authors provides, for the first time, a dynamic programming algorithm for the online monitoring of the robustness metric of MTL formulas with bounded future and unbounded past.
The past formula is used  to reason about the robustness of the actual system observations, while for evaluating the future formula they use a predictor to estimate the likely robustness. However, the value forecasted by the predictor needs to be trusted because it is not the real value that the system will provide.  Other approaches~\cite{JaksicBGKNN15,JaksicBGN18,rtamt} address the problem of deciding about the future using a technique called \emph{pastification} that rewrites the future operators as past ones and delays the verdict.  Similarly the works of Mamouras et al~\cite{Mamouras2020,MamourasCW21} delay 
the output verdict until some part of the future input is seen.  In~\cite{Deshmukh2017} the authors present an efficient online algorithm to compute the robust interval semantics for bounded horizon formulas. 
All these approaches 
assume that
the data and
the events to be observed come synchronously and in-order.

\noindent{\bf Our contribution}
In contrast to these works, we present a novel approach to monitor online imprecise spatio-temporal signals (signals are defined on intervals, not just the robustness) where the samples can be processed also out-of-order.  The notion of an interval is instrumental when representing partial knowledge about a value that is at least known to be within some boundaries. This might be because of errors in the measurement, or maybe because of some other sources of uncertainty throughout the process of acquiring and processing them.
We define both a Boolean and a quantitative interval semantics for STREL and we prove the soundness and the correctness of the robust interval semantics. We design and implement, as extensions to the Moonlight tool\footnote{Source code available at: \href{https://github.com/MoonLightSuite/MoonLight}{\texttt{github.com/MoonLightSuite/MoonLight}}}, the first online monitoring algorithm for not-in-order sampled signals, and the first online spatio-temporal monitoring tool. Our experiments demonstrate also convincing performances comparing with the state-of-the-art tool \textsc{Breach}~\cite{breach} for the online monitoring of temporal properties over in-order sampled signals. 

\noindent{\bf Paper organization} The rest of this paper is structured as follows.  We provide the important aspects of interval algebra and our notion of imprecise signals in Section~\ref{sec:background}. In Section~\ref{sec:logic} we introduce the interval extension of the STREL logic, and its primary results, while in Section~\ref{sec:monitoring} we present our approach for the online monitoring of imprecise signals.  Lastly, we present a realistic use case in Section~\ref{sec:evaluation} and we share our concluding remarks in Section~\ref{sec:future}.


%% file: sections/background.tex
\section{Interval Algebra, Signals and Spatial Model}
\label{sec:background}
In this section, we define the key elements of interval algebra, signal and spatial model, which will be useful to characterize samples of the kind depicted in Figure~\ref{fig:no2}.

\begin{definition}[Intervals]
Let $\mathcal{I}(\mathbb{R}^\infty)$ be the set of intervals defined over the set $\mathbb{R}^\infty \equiv \mathbb{R}\cup\{+\infty, -\infty\}$.
We call \emph{closed interval} (or simply \emph{interval}) any set $I\subseteq \mathbb{R}^\infty$ such that $I \equiv [a, b] := \{x \in \mathbb{R^\infty}: a \leqslant x \leqslant b; a, b \in \mathbb{R}^\infty\}$. 
For any $I \equiv [a, b] \in \mathcal{I}(\mathbb{R}^\infty)$, we will indicate as $\underline{I} \equiv a$ and $\overline{I} \equiv b$ the extremes of the interval.
\end{definition}

In addition to the classical notion of interval, it is useful to recall some basic operations that can be performed.

\begin{definition}[Interval Basic Operations]\label{def:interval-operations}
Consider $I, I_1, I_2 \in \mathcal{I}(\mathbb{R}^\infty), c \in \mathbb{R}^\infty$, we define the following interval operators:
\begin{align*}
c + I &:= [\underline{I}+c, \overline{I}+c]  
    &  -I &:= [-\overline{I}, -\underline{I}] \\
I_1 + I_2 &:= [\underline{I_1} + \underline{I_2}, \overline{I_1} + \overline{I_2}]  
    &  I_1 - I_2 &:= I_1 + (-I_2)
\end{align*}
\begin{center}
    $[\max](I_1, I_2) := [\max(\underline{I_1}, \underline{I_2}), \max(\overline{I_1}, \overline{I_2})]$ \\
$[\min](I_1, I_2) := [\min(\underline{I_1}, \underline{I_2}), \min(\overline{I_1}, \overline{I_2})]$
\\
\end{center}
We also consider the extensions of $[\min]$ and $[\max]$ operators defined over an arbitrary subset $A \subseteq \mathcal{I}(\mathbb{R}^\infty)$, denoted by $\{\}$ instead of $()$ for function arguments.\\
We call \emph{interval radius} of I, the operator  $|I|:= [\min(|\underline{I}|, |\overline{I}|), \max(|\underline{I}|, |\overline{I}|)]$
\end{definition}
Interval relations are described in this way:
\begin{definition}[Interval Inequalities]
Let $I_1, I_2 \in \mathcal{I}(\mathbb{R}^\infty)$, we say that $I_1 < I_2$ when $\overline{I_1} < \underline{I_2}$. Symmetrically, we say that $I_1 > I_2$ when $\underline{I_1} > \overline{I_2}$\footnote{ We will write $I < c$ (respectively $I > c$) in place of $I < [c, c]$ (resp. $I > [c, c]$)}, for $c \in \mathbb{R}^\infty$.
\end{definition}



To measure distances between intervals we consider the \emph{Hausdorff Distance.}
\begin{definition}[Hausdorff Distance]\label{def:hausdorff}
Let $X, Y$ be two non-empty subsets of a metric space $\langle M, d \rangle$, we will call (Hausdorff) distance the function $d_H : \mathcal{P}(X) \times \mathcal{P}(X) \rightarrow \mathbb{R}_{\geq 0}$ defined as
\begin{equation*}
    d_H := \max\left\{\,\sup_{x \in X} \inf_{y \in Y} d(x,y),\, \sup_{y \in Y} \inf_{x \in X} d(x,y)\,\right\}
\end{equation*}
 In practice, in our context, we can just consider the metric space defined by the euclidean distance over the real numbers, and thus $d_H$ reduces to computing $\max(|\underline{I_1} - \underline{I_2}|, |\overline{I_1} - \overline{I_2}|)$ for any two $I_1, I_2 \in \mathcal{I}(\mathbb{R}^\infty)$, although for doing that we say, by definition, that if both $\underline{I_1}, \underline{I_2}$ (or both $\overline{I_1}, \overline{I_2}$) are infinite, then their Hausdorff distance is $0$.
\end{definition}


Now we have all the tools to introduce the concept of imprecise signals.

\begin{definition}[Imprecise Temporal Signal]
Let $\mathbb{T} \equiv [0, \infty]$ be a set representing the time domain, and let $\mathcal{F}(\mathbb{T}, D^n)$, with $D \subseteq \mathcal{I}(\mathbb{R})$ for a fixed $n \in \mathbb{N}$, be the family of functions over Cartesian products of real intervals; we call \textit{imprecise time signal}, any $\mathbf{\sigma} \in \mathcal{F}(\mathbb{T}, D^n)$, i.e. any function $\mathbf{\sigma}: \mathbb{T} \rightarrow D^n$
\end{definition}

It is convenient, in some cases, to slice the signals based on the domain $D$ of interest, for that reason, we recall the concept of (signal) projection.

\begin{definition}[Signal Projection]\label{def:projection}
 Let $\pi_i:D_1 \times \dots \times D_i \times \dots \times D_n \rightarrow D_i$ be the function that  takes the $i$-th projection of the set-theoretic Cartesian product $D^n$, we will indicate as $\pi_i(\mathbf{\sigma}(t))$ the projection of $\mathbf{\sigma}(t)$ to the $i$-th 1-dimensional signal $s_i:\mathbb{T} \rightarrow D_i$.
\end{definition}

To represent a set of signals distributed in the space, we introduce the following definition.
\begin{definition}[Imprecise Spatio-Temporal Signal]\label{def:signal}
Let $\mathcal{F}(\mathbb{L}, \mathbb{T}, D^n)$ be the family of functions of space and time over real intervals, with $\mathbb{L}$ a set of locations,
we call \textit{imprecise spatio-temporal signal} -- or just \textit{signal} when there is no risk of ambiguity -- any $s \in \mathcal{F}(\mathbb{L}, \mathbb{T}, D^n)$, i.e. any function:
$    \mathbf{s}: \mathbb{L} \times \mathbb{T} \rightarrow D^n$
\end{definition}

Considering the pollution example, where $\mathbb{L}$ is the set of stations, $\mathbb{T}=[0,10]$ is the time domain corresponding to an interval of 10 days, and $D$ is the possible range for nitrogen-dioxide values  ($NO_2$) in the air; then the spatio-temporal signal $\mathbf{s}: \mathbb{L} \times \mathbb{T} \rightarrow D$ returns at each time, in each location the value of $NO_2$, $\mathbf{s}(\ell, t)$.

We can naturally describe the distance between spatio-temporal signals by considering the Hausdorff distance from Definition~\ref{def:hausdorff} over all possible locations of the space, and time instants.

\begin{definition}[Spatio-Temporal Signal Distance]\label{def:signal-distance}
Let $\mathbf{s_1}, \mathbf{s_2} \in \mathcal{F}(\mathbb{L}, \mathbb{T}, D^n)$, we will call \emph{signal distance} the largest Hausdorff distance over space and time, defined as:
\begin{equation*}
    ||\mathbf{s_1}-\mathbf{s_2}||_\infty := \max\limits_{i \leq n}\max\limits_{\ell \in \mathbb{L}}\max\limits_{t \in \mathbb{T}}\{d_H(\pi_i(\mathbf{s_1}(\ell, t)), \pi_i(\mathbf{s_2}(\ell, t)))\}
\end{equation*}
\end{definition}

To describe the interplay of signals in different locations, we need to encompass the information related to the spatial distribution of the locations.

\begin{definition}[Spatial model]
We call \emph{spatial model} the tuple $\mathcal{S} = \langle \mathbb{L}, W \rangle$ 
\footnote{We focus on real-valued positive labels, to convey the intuitive meaning of distance between two locations. For alternative definitions of $W$, the interested reader might refer to~\cite{Bartocci17memocode}.}
, where $\mathbb{L}$ is a set of locations
and $W \subseteq \mathbb{L} \times  \mathbb{R}_{\geq 0}^{\infty} \times \mathbb{L}$ is a \emph{proximity function} associating at most one label $w \in \mathbb{R}_{\geq 0}^{\infty}$ to each distinct pair $\ell_1, \ell_2 \in \mathbb{L}$
\end{definition}

An obvious spatial model for the region of Figure~\ref{fig:no2} is a graph where every location is connected to all the others, and the proximity function is defined by labels corresponding to the minimal aerial distance between each pair of locations. Finally to consider distance on paths of locations we introduce the notion of \textit{routes} over the spatial model.

\begin{definition}[Routes]
A route $\tau$ on $\mathcal{S}$ is a (potentially infinite) sequence $\ell_0, \ell_1, \dots \ell_k \dots $, such that for any $\ell_i,\ell_{i+1} \in \tau$, there is a label $(\ell_i, w, \ell_{i+1}) \in W$.
We indicate by $\Lambda(\ell)$ the set of routes on $\mathcal{S}$ starting at $\ell$. Moreover, we will use $\tau[i]$ to denote the $i$-th node of the route, $\tau[i\dots]$ to denote the subroute starting at the $i$-th node, and $\tau(\ell_i)$ to denote the first occurrence of $\ell_i$ in $\tau$. Lastly, we will indicate by $\ell_1<\tau(\ell_2)$ the fact that $\ell_1$ precedes $\ell_2$ in the route $\tau$.
\end{definition}

Routes have the same intuitive meaning as they have in the physical world, and, similarly to the real world, we can define the concept of route (or travel) distance, as the aggregated sum of all the labels traversed by the route.

\begin{definition}[Route Distance]
For a given $\tau$ on $\mathcal{S}$, the distance $d_\tau[i]$ is:
\begin{equation*}
    d_\tau[i] = \begin{cases} 
                0, & \mbox{if } i = 0 \\ 
                w + d_{\tau[1\dots]}[i - 1], & \mbox{if } i > 0 \mbox{ and } (\tau[0], w, \tau[1]) \in W
               \end{cases}
\end{equation*}
\end{definition}

Lastly, routes allow us to conveniently define the distance between any two locations $\ell_1, \ell_2 \in \mathbb{L}$, whichever the spatial model being considered. In fact, from the location $\ell_1$ to $\ell_2$, one can consider the minimal distance among all the routes starting at $\ell_1$ and ending in $\ell_2$: 
\begin{equation*}
    d_\mathcal{S}[\ell_1, \ell_2] = \min\limits_{\tau \in \Lambda(\ell_1)}\{d_\tau[\ell_2]\}
\end{equation*}


%% file: sections/logic.tex
\section{STREL with interval Semantics}\label{sec:logic}
We present in this section an interval semantics that allows for a conservative analysis that considers both the minimum and the maximum values of intervals. This way, a plethora of use scenarios can fit into this specification language, spanning from traditional offline monitoring of a given specification over imprecise signals to online monitoring with out-of-order updates. All the proofs of theorems and lemmas are reported in appendix.

\begin{definition}[STREL Syntax]
We consider logical formulae belonging to the language $\mathcal{L}$ generated by the following BNF grammar:
\begin{equation*}
    \varphi := \top~|~\bot~|~p~\circ~c~|~\neg~\varphi~|~\varphi~\lor~\varphi~|~\varphi~\until{I}~\varphi
    ~|~\varphi~\reach{\leq d}~\varphi~|~\escape{\geq d}~\varphi
\end{equation*}
where $\circ \in \{>, <\}$, $c \in \mathbb{R}$, $p$ is associated to a projection function $\pi$ of Definition~\ref{def:projection}, i.e. $p~\circ~c$ are inequalities on the variables of the systems. $\until{I}$ is the  \emph{until} temporal operator, with $I$ real interval, while  $\reach{\leq d}{}$ and $\escape{\geq d}{}$ are the spatial operators \emph{reach} and \emph{escape} , with $d \in  \mathbb{R}_{>0}$.
In addition, we have the derived Boolean operators as \emph{and} ($\land$) and \emph{implies} ($\to$), temporal operators \emph{eventually} ($\ev{I}$) and \emph{globally} ($\glob{I}$), and spatial operators  \emph{somewhere} ($\somewhere{\leq d}{}$) and \emph{everywhere} ($\everywhere{\leq d}{}$).
\end{definition}
Considering  again the air pollution case study, 
current regulation in Lombardy requires to take action when the level of nitrogen dioxide ($NO_2$) exceeds the threshold of $400\mu g/m^3$ for more than three hours. Let $NO_2 < 400$ denote the atomic proposition that states that the level of nitrogen dioxide is lower than $400\mu g/m^3$. A requirement as the previous one could be expressed like in~(\ref{p1}):
\begin{equation}\label{p1}
    \ev{[0, 3hours]} NO_2 < 400
\end{equation}
Temporal operators like $\ev{}$ specify properties on the dynamic evolution of the system. In fact, when (1) is violated, the alerting procedure could be triggered, to inform the citizens about the danger. 
However, since it is known that noise and local faults frequently happen, one could consider of alerting the population also when the close neighbourhood (e.g. within 10 Km) exhibits a similar phenomenon. For this aim, a property like (\ref{p2}) can be monitored.
\begin{equation}\label{p2}
    \somewhere{<10km}{} NO_2 < 400
\end{equation}

Spatial operators like $\somewhere{}{}$ instead specify properties related to the spatial configuration, and in this context, 
the exact meaning is that at least a location in a range of less than 10 km must have a level of nitrogen dioxide lower than $400\mu g/m^3$.
We will see other examples of the logic language in the next sections. For a more detailed description of the logic, we refer the reader to~\cite{Bartocci17memocode}.
We present now the Boolean and quantitative interval semantics for STREL.
\begin{definition}[STREL Boolean Semantics]\label{def:boolean-semantics}
Let $\chi:\mathcal{F}(\mathbb{L}, \mathbb{T}, D^n)$ $ \times \mathbb{L} \times \mathbb{T} \times \mathcal{L} \rightarrow \{-1, 0,1\}$ be a function defined as follows:
\begin{itemize}[leftmargin=-.03in]
    \setlength\itemsep{.5em}
    \item[] $\chi(\mathbf{s}, \ell, t, \top) = 1$
    \item[] $\chi(\mathbf{s}, \ell, t, \bot) = -1$
    \item[] $\chi(\mathbf{s}, \ell, t, p \circ c) = 
            \begin{cases} 
                1, & \mbox{if } \pi_p(\mathbf{s}(\ell,t)) \circ c \\ 
                -1, & \mbox{if } \pi_p(\mathbf{s}(\ell,t)) \circ^{-1} c \\ 
                0, & \mbox{otherwise}\footnotemark
            \end{cases}$
    \item[] $\chi(\mathbf{s}, \ell, t, \lnot \varphi) = - \chi(\mathbf{s}, \ell, t, \varphi)$
    \item[] $\chi(\mathbf{s}, \ell, t, \varphi_1 \lor \varphi_2) = \max(\chi(\mathbf{s}, \ell, t, \varphi_1), \chi(\mathbf{s}, \ell, t, \varphi_2))$
    \item[] $\chi(\mathbf{s}, \ell, t, \varphi_1 \until{I} \varphi_2) = \max\limits_{t'\in t + I}\{\min(\chi(\mathbf{s}, \ell, t', \varphi_2), \min\limits_{t''\in [t', t]}\{\chi(\mathbf{s}, \ell, t'', \varphi_1)\})\}$
    \item[] $\chi(\mathbf{s}, \ell, t, \varphi_1 \reach{\leq d}{} \varphi_2) = \max\limits_{\tau \in \Lambda(\ell)}\max\limits_{\ell' \in \tau : d_\tau[\ell'] \leq d}\\[3pt]  
    \null\qquad\qquad\qquad\qquad\qquad~~\{\min(\chi(\mathbf{s}, \ell', t, \varphi_2),\min\limits_{\ell'' < \tau(\ell')}\{\chi(\mathbf{s}, \ell'', t, \varphi_1)\}\}$
    \item[] $\chi(\mathbf{s}, \ell, t, \escape{\geq d}{} \varphi) = \max\limits_{\tau \in \Lambda(\ell)} \max\limits_{\ell' \in \tau : d_\mathcal{S}[\ell,\ell'] \geq d}
                \min\limits_{\ell'' < \tau(\ell')}\{\chi(\mathbf{s}, \ell'', t, \varphi)\}$
\end{itemize}
\end{definition}
\footnotetext{Note that `$>$' and `$<$'are used in this context to represent interval inequalities, which do not define a total ordering.} 
This is a three-valued semantics, which is equal to $1$ if the interval signal $\chi(\mathbf{s}, \ell, t, \varphi)$ satisfies $\varphi$, $-1$ if the formula is not satisfied, and 0 if we cannot answer. The semantics is directly derived from the standard Boolean semantics and the interval algebra described in the previous section. For atomic proposition $\chi(\mathbf{s}, \ell, t, p \circ c)=1$ iff the inequality $\pi_p(\mathbf{s}(\ell,t))\circ c$ is true.
This means, e.g.,  $\chi(\mathbf{s}, \ell, t, NO_2>400)=1$ iff $\underline{NO_2}$, the left extreme of the projected signal $NO_2$,  is greater than 400. $\chi(\mathbf{s}, \ell, t, NO_2>400)=-1$ if the right extreme is less than 400, and $\chi(\mathbf{s}, \ell, t, NO_2>400)=0$ otherwise, so if $400 \in NO_2$ interval value. Similar calculation can be done for the other combination of $\circ$ and $c$.

The three-valued Boolean semantics can be sufficient in applications where the interest is only whether or not the specification is satisfied.   However, in many complex cases, one might be interested in getting some insights about the degree by which a property is satisfied or violated. In the following, we introduce an extension of the quantitative semantics that provides numerical bounds to the robustness degree of a specification.

\begin{definition}[STREL Robust Interval Semantics]\label{def:interval-semantics}
Let $\rho:\mathcal{F}(\mathbb{L}, \mathbb{T}, D^n) \times \mathbb{L} \times \mathbb{T} \times \mathcal{L} \rightarrow \mathcal{I}(\mathbb{R})$ be the function mapping signals, locations, time instants, and formulae defined as follows:
\begin{itemize}[leftmargin=-.1in]
    \setlength\itemsep{.8em} 
    \item[] $\rho(\mathbf{s}, \ell, t, \top) = [+\infty, +\infty]$
    \item[] $\rho(\mathbf{s}, \ell, t, \bot) = [-\infty, -\infty]$
    \item[] $\rho(\mathbf{s}, \ell, t, p \circ c) = 
        \begin{cases} 
                \pi_p(\mathbf{s}(\ell,t)) - c, & \mbox{if } \circ \mbox{ is } ` >  \text{'}\\ 
                c - \pi_p(\mathbf{s}(\ell,t)), & \mbox{if } \circ \mbox{ is } ` <  \text{'}
            \end{cases}$
    \item[] $\rho(\mathbf{s}, \ell, t, \lnot \varphi) = - \rho(\mathbf{s}, \ell, t, \varphi)$
    \item[] $\rho(\mathbf{s}, \ell, t, \varphi_1 \lor \varphi_2) = [\max](\rho(\mathbf{s}, \ell, t, \varphi_1), \rho(\mathbf{s}, \ell, t, \varphi_2))$
    \item[] $\rho(\mathbf{s}, \ell, t, \varphi_1 \until{I} \varphi_2) = \left[\max\limits_{t'\in t + I}\right] \\[4pt]  
    \null\qquad\qquad~~\left\{[\min]\left(\rho(\mathbf{s}, \ell, t', \varphi_2),
    \left[\min\limits_{t''\in [t', t]}\right]\{\rho(\mathbf{s}, \ell, t'', \varphi_1)\}\right)\right\}$
    \item[] $\rho(\mathbf{s}, \ell, t, \varphi_1 \reach{\leq d}{} \varphi_2) = \left[\max\limits_{\tau \in \Lambda(\ell)}\right] \left[\max\limits_{\ell' \in \tau : d_\tau[\ell'] \leq d}\right]\\[4pt]  
    \null\qquad\qquad~~\left\{[\min](\rho(\mathbf{s}, \ell', t, \varphi_2),\left[\min\limits_{\ell'' < \tau(\ell')}\right]\{\rho(\mathbf{s}, \ell'', t, \varphi_1)\}\right\}$
    \item[] $\rho(\mathbf{s}, \ell, t, \escape{\geq d}{} \varphi) = \left[\max\limits_{\tau \in \Lambda(\ell)}\right] \left[\max\limits_{\ell' \in \tau : d_\mathcal{S}[\ell, \ell'] \geq d}\right]\left[\min\limits_{\ell'' < \tau(\ell')}\right] \\[4pt] 
    \null\qquad\qquad\qquad\qquad~~\{\rho(\mathbf{s}, \ell'', t, \varphi)\}$
\end{itemize}
\end{definition}

We will indicate with $\rho^{\varphi}_{s}: \mathbb{L} \times \mathbb{T} \rightarrow \mathcal{I}(\mathbb{R}^\infty)$ the \emph{robustness signal}, i.e. the signal generated by the partial application of the $\rho$ function to a given formula $\varphi$ and a given signal $\mathbf{s}$, so that $\rho^{\varphi}_{s}(\ell, t) \equiv \rho(\mathbf{s}, \ell, t, \varphi)$.   

Note that without the interval semantics we have defined, missing values should be substituted by some values that approximate the actual value (e.g. by linear interpolation), and therefore only approximate the actual value of satisfaction or robustness of a given property at that specific time point. Conversely, by exploiting the interval semantics, one could actually get upper/lower bounds at those points, which can actually be sufficient in real-world applications.

\begin{theorem}[Soundness of Robust Interval Semantics]\label{th:soundness}
The Robust Interval Semantics of Definition~\ref{def:interval-semantics} is sound w.r.t the Boolean Semantics of Definition~\ref{def:boolean-semantics}, i.e. 
for any $\mathbf{s} \in \mathcal{F}(\mathbb{L}, \mathbb{T}, D^n)$, $\ell \in \mathbb{L}$, $t \in \mathbb{T}$, $\varphi \in \mathcal{L}$:
\begin{itemize}
    \item[] $ \mbox{if\quad} \rho(\mathbf{s},\ell, t, \varphi) > 0 \mbox{\quad then \quad} \chi(\mathbf{s}, \ell, t, \varphi) = 1 $
    \item[] $ \mbox{if\quad} \rho(\mathbf{s},\ell, t, \varphi) < 0 \mbox{\quad then \quad} \chi(\mathbf{s}, \ell, t, \varphi) = -1$
    \item[] $ \mbox{if\quad} 0 \in \rho(\mathbf{s},\ell, t, \varphi)   \mbox{\quad then \quad} \chi(\mathbf{s}, \ell, t, \varphi) = 0$
\end{itemize}
\end{theorem}
\begin{proof}
See the extended version of this article for the proof. 
\end{proof}

To provide the correctness of the interval semantics over imprecise signals, we introduce the following lemma:
\begin{lemma}[Metric Lemma]\label{th:metric-lemma}
Let $\mathbf{s_1}, \mathbf{s_2} \in \mathcal{F}(\mathbb{T}, D^n)$. For any $t \in \mathbb{T}$, for any $\ell \in \mathbb{L}$, for any $\varphi \in \mathcal{L}$, for any $\delta > 0$,  we have:
\begin{equation*}
   \mbox{if\quad}||\mathbf{s_1}-\mathbf{s_2}||_\infty < \delta \mbox{\quad then \quad} ||\rho_{s_1}^{\varphi}-\rho_{s_2}^{\varphi}||_\infty < \delta
\end{equation*}
\end{lemma}
\begin{proof}
See the extended version of this article for the proof. 
\end{proof}

\begin{theorem}[Correctness of Robust Interval Semantics]\label{th:correctness}
The Robust Interval Semantics of Definition~\ref{def:interval-semantics} is \emph{correct} w.r.t the Boolean Semantics of Definition~\ref{def:boolean-semantics}, i.e. 
for any $\mathbf{s} \in \mathcal{F}(\mathbb{L}, \mathbb{T}, D^n)$, $\ell \in \mathbb{L}$, $t \in \mathbb{T}$, $\varphi \in \mathcal{L}$:
\begin{equation*}
    \mbox{if\hspace{0.5em}} ||\mathbf{s_1}-\mathbf{s_2}||_\infty < | \rho(\mathbf{s}_1,\ell, t,\varphi)| \mbox{\hspace{0.5em}then\hspace{0.5em}} \chi(\mathbf{s}_1, \ell, t,\varphi) = \chi(\mathbf{s}_2, \ell, t,\varphi)
\end{equation*}
 for all $i \leq n$, where $|\cdot|$ is the interval radius of Definition~\ref{def:interval-operations}. 
\end{theorem}
\begin{proof}
See the extended version of this article for the proof. 
\end{proof}

%% file: sections/monitoring.tex
\section{Online Monitoring}\label{sec:monitoring}

In this section a novel \emph{online (out-of-order) monitoring algorithm} for STREL is presented. Differently from the standard \emph{offline approach}, where all the data is available at the beginning of the execution, \emph{online} monitoring is performed incrementally, when a new piece of data is available. In this case, the uncertainty related to the absence of information must be taken into account.
For that aim, the machinery of imprecise signals can be exploited to represent the uncertainty, where the result of the monitoring process, whether it is a satisfaction or a robustness signal, is refined as soon as new updates of the input arrive. 

\begin{figure}[htpb]
  \centering
  \includegraphics[width=1\columnwidth]{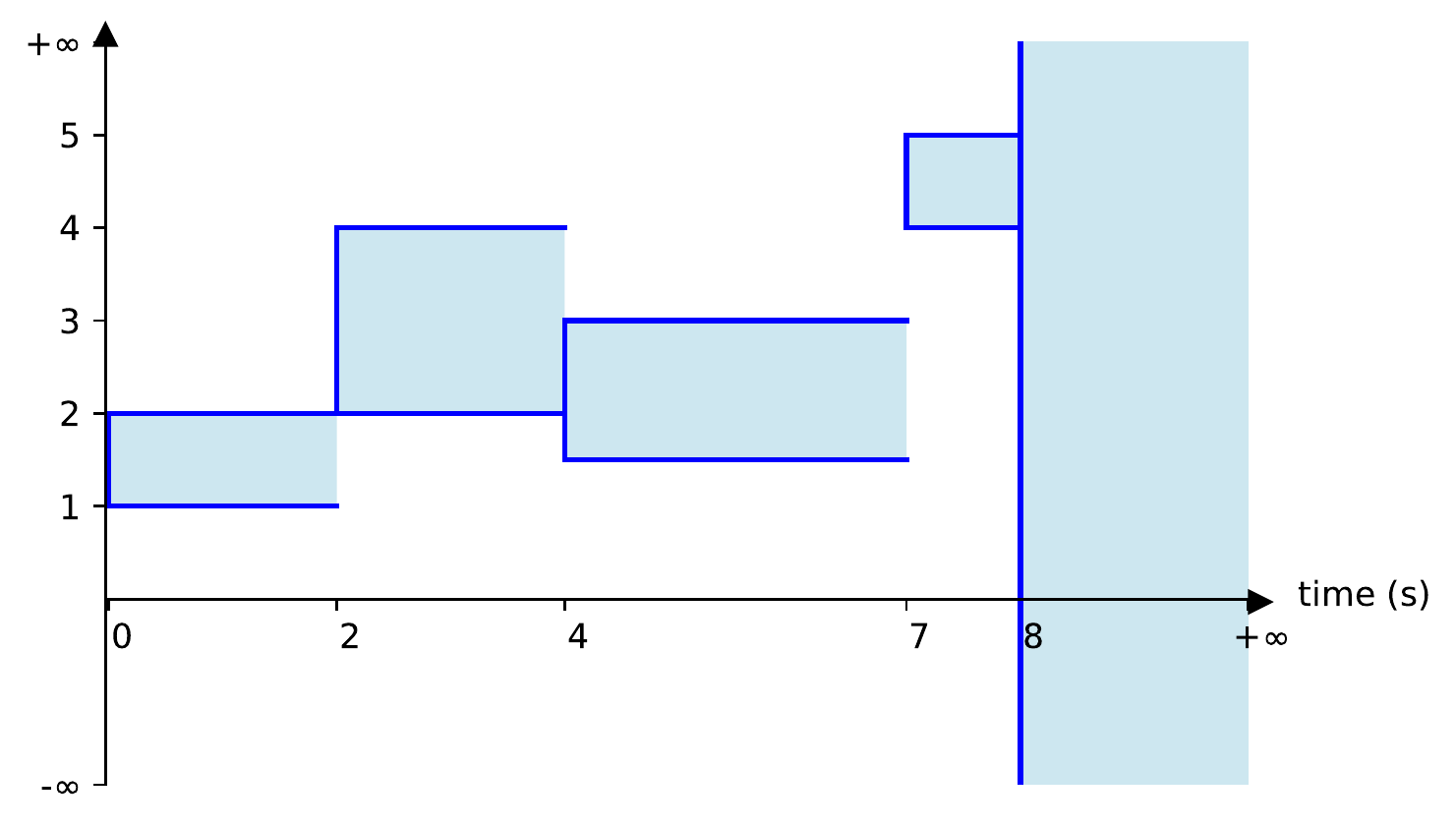}
  \caption {
    A graphical view of a typical piecewise-constant imprecise signal. From the 8-th second onward, there is total absence of knowledge about what values the signal could have, but until then, we have some bounds on the actual values observed.
  }
  \label{fig:signal-appearance}
\end{figure}

The semantics for STREL is defined for arbitrary signals, but algorithms, for computational reason, are provided for piecewise constant ones, along the lines of~\cite{Deshmukh2017,Bartocci17memocode}. This class of signals is convenient, and frequently chosen as the class of reference for a number of reasons: (i) it naturally describes digital signals, (ii) it can be stored in memory very efficiently, and processed fast enough to be considered for real-time applications, (iii) it allows to express the vast majority of real-valued signals of practical use with a limited loss of information. 
Since the presented signals are Lipschitz-continuous (we consider only inequalities on the variables of the system), we can always bound our error, considering the minimum time step and the maximum of their individual Lipschitz constants.
An \emph{imprecise piecewise-constant signal} $\sigma: \mathbb{T} \rightarrow  \mathcal{I}(\mathbb{R}^\infty)$, can be characterized in the following way:

\[
\sigma(t) = 
\begin{cases}  
        I_{i}, &  \textrm{for}~t_i \leq t < t_{i+1}\\
        \vdots \\
        I_{n}, &  \textrm{for}~t_{n} \leq t < \infty
\end{cases}
\]

and graphically represented as in Figure~\ref{fig:signal-appearance}. Note that frequently when monitoring real-time application the last part of the signal will be characterized by the widest interval possible, as this denotes the fact that the knowledge collected so far is insufficient for providing any insight about the monitored specification for future values of the signal. Similar infinite interval can be considered for missed values.



We consider \emph{space-synchronized} (s.s.) signals, i.e. signals defined on the same time intervals for any location of the space model.
More precisely, a p.c. s.s. signal $s: \mathbb{L} \times \mathbb{T} \rightarrow D^n$ is a signal that can be represented as a sequence of pairs $\{(t_i, \mathbf{V}_i)\}_{i \in \mathbb{N}}$, where each pair $(t_i, \mathbf{V}_i)$ of the sequence represents a piece of the signal, such that it maps any time-instant between $t_i$ and $t_{i+1}$ in $\mathbb{T}$, 
to the $|\mathbb{L}| \times n$ matrix $\mathbf{V}_i$ that represents the values of the $n$ dimensions of the signal at each location $\ell$ in $\mathbb{L}$.
The space-synchronization restriction might appear to be a severe limitation, but this shows one of the conceptual differences between online and offline monitoring: in an offline setting, the space-synchronization hypothesis would likely have detrimental effects on the performances, as it would force all the processing to happen at a temporal granularity that is the union of the temporal granularities of the signals at the different locations. In an online setting, on the other hand, the temporal granularity is determined by the time when new information is available, and the space-synchronization hypothesis makes it possible to exploit in future work the Single-Instruction Multiple-Data (SIMD) capabilities of modern processors (see~\cite{Kusswurm2020, hayes2016}), resulting in execution times that are virtually independent from the number of locations, when appropriate hardware is available.
In this context, we call \emph{signal update} $\mathbf{u}$ the triplet $(t_a, t_b, \mathbf{V})$, representing a mapping to the value matrix $\mathbf{V}$ for any time instant between $t_a$ (included) and $t_b$ (excluded). 
Signal updates can be seen as some special kinds of signals that we use to represent upcoming partial information from the online behavior of the monitored system.
In the context of our analysis, we always assume updates to provide truthful information (the case of hard faults, i.e. where updates provide wrong information, will be explored in future work), and, for that reason, we can always assume updates to be well-formed, meaning that the interval $\mathbf{V}$ they provide is always included in the previous interval of the signal we stored for that time and location.
To express the \emph{online} nature of the computation we want to pursue, we need some way of describing the incremental evaluation of new information.

\begin{definition}[Signal refinement]
Let $\mathbf{s_1}, \mathbf{s_2} \in \mathcal{F}(\mathbb{L}, \mathbb{T}, D^n)$, 
we say that $\mathbf{s_1}$ \textit{is refined by} $\mathbf{s_2}$, 
and we write $\mathbf{s_1} \succ \mathbf{s_2}$, 
iff for any $\ell \in \mathbb{L}, t \in \mathbb{T}, i \leq n$,  $\pi_i(\mathbf{s_2}(\ell, t)) \subseteq \pi_i(\mathbf{s_1}(\ell, t))$, 
and there is some $\ell' \in \mathbb{L}, t' \in \mathbb{T}, i' \leq n$, 
such that $\pi_{i'}(\mathbf{s_2}(\ell', t')) \subset \pi_{i'}(\mathbf{s_1}(\ell', t'))$,
i.e. each interval of the co-domain of the signal $\mathbf{s_2}$ is contained in the corresponding interval of the signal $\mathbf{s_1}$, and some of them are strictly contained. 
\end{definition}

The \textit{refinement} relation expresses the fact that $\mathbf{s_1}$ and $\mathbf{s_2}$ represent the same information, except that $\mathbf{s_2}$ has a smaller degree of uncertainty.
By the notions of signal update and signal refinement, we can easily represent the online evolution of a signal as a chain of signal refinements $\mathbf{s_0} \succ \dots \succ \mathbf{s}_j \succ \dots $, where the signal $\mathbf{s}_{j+1}$ at the step $j+1$ can be computed from $\mathbf{s}_{j}$ and update $\mathbf{u}_{j}$ like in Algorithm~\ref{alg:refinement}.

\begin{algorithm}
    \caption{Signal Refinement} \label{alg:refinement}
    \begin{algorithmic}[1]
       \Procedure{refine}{\scalebox{0.9}{$\mathbf{s}:\{(t_0,\mathbf{V}_0),\dots,(t_N,\mathbf{V}_N)\}$, $\mathbf{u}:(t_a, t_b, \mathbf{V})$}}
        \For{$(t_i, \mathbf{V}_i)$ in $\mathbf{s}$}
            \If{$t_i < t_a <t_{i+1}$ \textbf{and} $\mathbf{V} \subset \mathbf{V}_i$}
                \State $\mathbf{s}$ := $\mathbf{s} \cup (t_a, \mathbf{V})$ 
            \ElsIf{$t_i = t_a$  \textbf{and} $\mathbf{V} \subset \mathbf{V}_i$}
                \State $\mathbf{s}$ := $\mathbf{s} \setminus (t_i, \mathbf{V}_i) \cup (t_i, \mathbf{V})$ 
            \ElsIf{$t_a < t_i < t_b$}
                \State $\mathbf{s}$ := $\mathbf{s} \setminus (t_i, \mathbf{V}_i)$ 
                
            \EndIf
            \If{$t_i < t_b < t_{i+1} $}
                    \State $\mathbf{s}$ := $\mathbf{s} \cup (t_b, \mathbf{V}_i)$ 
            \EndIf
        \EndFor
        \EndProcedure 
    \end{algorithmic}
\end{algorithm}
The $\Call{refine}{~}$ procedure takes a signal $\mathbf{s}_j$ as a sequence of ordered pairs, and an update as the triplet $(t_a, t_b, \mathbf{V})$. In practice, it removes all the pieces of the signal that start within the interval $[t_a, t_b)$, and adds a piece with value $\mathbf{V}$  in the case $t_a$ and/or  $t_b$  lay in between of $t_i$ and $t_{i+1}$. Clearly, for efficiency reasons, the algorithm can jump to the next pair each time $t_{i+1} < t_a$, and can terminate as soon as $t_i > t_b$.
The updated signal at the end of the execution is the next element of the refinement chain, i.e. $\mathbf{s}_{j+1}$.

\noindent{\bf The Monitoring Problem.}
When monitoring online a given specification $\varphi$, let $\mathbf{s}_0$ be the signal representing the starting information on which the atoms of the formula $\varphi$ are defined. Let also $(\mathbf{u}_j)_{j \in \mathbb{N}}$ denote a (finite or infinite) sequence of signal updates.
The online monitoring problem can be framed as computing the robustness signal $\rho_{\mathbf{s}_{j+1}}^{\varphi}$, given $\rho_{\mathbf{s}_j}^{\varphi}$ (or, alternatively, the satisfaction signal $\chi_{\mathbf{s}_{j+1}}^{\varphi}$ given $\chi_{\mathbf{s}_j}^{\varphi}$), with $\mathbf{s}_{j} \succ \mathbf{s}_{j+1}$, starting from $\mathbf{s}_0$.
%
A naïve implementation of an online monitor could just ignore the information coming from previous monitoring steps and restart the computation over the whole signal each time new information is available. As already noted in~\cite{online-fainekos,Deshmukh2017}, such an implementation would result in huge amounts of wasted resources when monitoring time signals, and it would therefore be even more costly when monitoring space-time signals.
To properly scope the effect that an out-of-order update of the input signal generates for the evaluation of a formula, it is convenient to think of updates as starting from the atoms of the monitored formula, and then propagating their effects up through the syntactic tree, generating a ripple effect where the impacted time span widens based on the operators of the subformulae. Figure~\ref{fig:time-ripple} shows the ripple effect resulting from the propagation of update information through the syntactic tree.
From this intuition we can define the \textit{update ripple} function, to scope the resulting update's time span, based on the provided update, and on the operator being computed in the following way:

\begin{figure}\hspace*{-0.35cm}   
    \includegraphics[width=1.1\columnwidth]{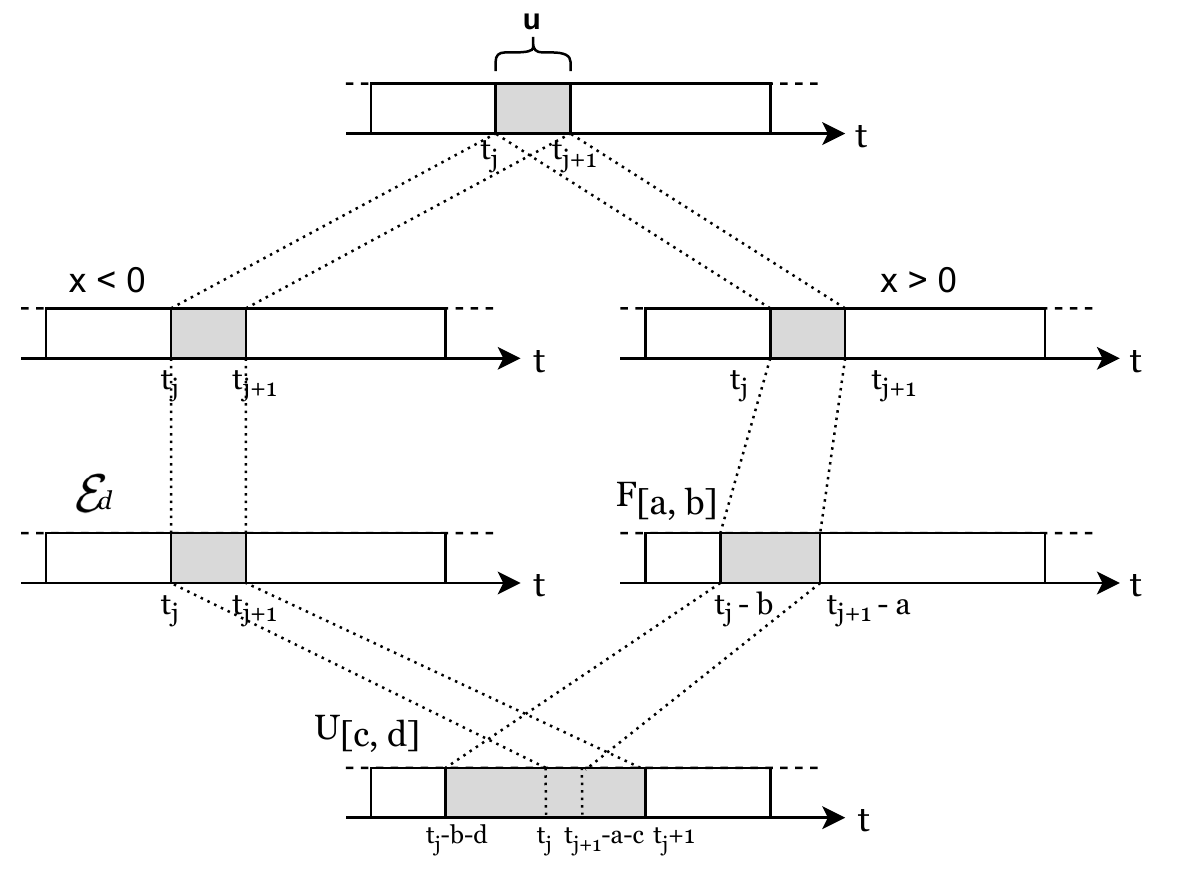}
    \caption{The ripple effect generated from the propagation of an update of the signal $x$, over the formula $\varphi :=  (\escape{d}{} x < 0) ~\until{[c,d]} (\ev{[a,b]} x > 0) $.}
    \label{fig:time-ripple}
\end{figure}


\begin{definition}[Update Ripple]\label{def:ripple}
Let $\mathbf{OP}_I$ denote any temporal operator on the time interval $I$, given a signal update $\mathbf{u} \equiv (t_a, t_b, \mathbf{V})$, we call \textit{update ripple} the function
\begin{equation*}
   \mathfrak{ur}(\mathbf{u}, \varphi) = 
   \begin{cases}
        [t_a, t_b) - I  & \varphi \equiv \mathbf{OP}_I \psi~\mathrm{or}~\varphi \equiv \psi_1 \mathbf{OP}_I \psi_2\\
        [t_a, t_b) & \mathrm{otherwise}
   \end{cases}
\end{equation*}
\end{definition}

Being able to assess the time boundaries of the effect of an update, we can therefore define an online monitoring procedure that updates the robustness (or satisfaction) signal when needed and that keeps the valid parts otherwise.
In general, we have that at the step $j+1$, the monitoring function can be evaluated as:
\begin{equation*}
   \rho_{j+1}(\mathbf{s}_{j+1}, \ell, t, \varphi) = 
   \begin{cases}
        \Call{monitor}{\mathbf{u}_{j}, \varphi}[\ell, t]  & t \in \mathfrak{ur}(\mathbf{u}_j, \varphi)\\
        \rho_{j}(\mathbf{s}_j, \ell, t, \varphi) & otherwise
   \end{cases}
\end{equation*}
        
Note that in all the cases where the updates overlap, they must be processed sequentially in order to generate correct results.


\noindent{\bf Monitoring Procedure.}
To compute the monitoring result signal online, it is crucial to be able to exploit the knowledge from the past each time new information is available. 
The most natural way for doing so is to develop a stateful algorithm that stores the relevant information from previous computations.
We will represent by $\mathcal{M}$ the persistent memory (i.e. the state) that we keep throughout the various iterations of the monitoring process, and by $\mathcal{M}[x]$ the access to the item $x$ from memory.
The memory $\mathcal{M}$ is organized around a data structure that represents the set of robustness signals of all the subformulae of the monitored formula $\varphi$, as computed in the last iterations. 
This set can be encoded as an array indexed on some ordering of the subformulae. We represent as $\mathcal{M}[\rho^\psi]$ the access to the respective robustness signal for some formula $\psi$.
This data structure is extremely important to maximize the time performance of the monitoring process, as next iterations will re-compute only the differing fragments based on the update ripple. 
Before starting the monitoring process, the memory is initialized by storing an \textit{undefined signal} for any subformula $\psi$ in the set of the subformulae of the formula $\varphi$ being monitored.
We call \emph{undefined signal} the special signal $s: \mathbb{L}\times \mathbb{T} \rightarrow [-\infty, +\infty]$, which represents the total absence of knowledge about the value, at any possible time instant.

Once the memory is initialized, the monitoring can start. We assume that the signal is always received as a sequence of signal updates $\mathbf{u}_j$, starting from $j=0$, where the input signal is considered to be undefined. Algorithm~\ref{alg:monitoring} represents the base routine triggered when receiving an update $\mathbf{u}$ of the input signal.
The recursive procedure \Call{monitor}{$\mathcal{M}, \varphi, \mathbf{u}$} is responsible for propagating the input signal update to the subformulae and then fetching the corresponding updates of the robustness signal. We indicate by $\langle \mathcal{M}, \{u^\varphi\} \rangle$ the return value of the algorithm, to mean that it returns an updated version of the memory, and a list of robustness updates of the formula $\varphi$ that might either be used by the caller or discarded. 
The general procedure of Algorithm~\ref{alg:monitoring} calls the specific procedures of Algorithms~\ref{alg:atoms}-\ref{alg:sliding-window} depending on the operators encountered while traversing the tree of the formula. Note that all of the above exploit the \Call{refine}{~} primitive operation from Algorithm~\ref{alg:refinement}.


\begin{algorithm}
    \caption{Online Monitoring Procedure} \label{alg:monitoring}
    \begin{algorithmic}[1]
        \Procedure{monitor}{$\mathcal{M}, \varphi, \mathbf{u}$}
        \Switch{$\varphi$}
            \Case{$p \circ c$}
                \State $\langle \mathcal{M}, \{u^\varphi\} \rangle$ :=  \Call{atom}{$\mathcal{M}, \varphi, \mathbf{u}$}
            \EndCase
            \Case{$\lnot \psi$ \textbf{or} $\escape{\geq d}{} \psi$}
                \State $\langle \mathcal{M}, \{u^\varphi\} \rangle$ :=  \Call{unary}{$\mathcal{M}, \varphi, \mathbf{u}$}
           \EndCase
           
            \Case{$\psi_1 \lor \psi_2$ \textbf{or} $\psi_1 \reach{\leq d}{} \psi_2$}
                \State $\langle \mathcal{M}, \{u^\varphi\} \rangle$ :=  \Call{binary}{$\mathcal{M}, \varphi, \mathbf{u}$}
            \EndCase
            
            \Case{$\ev{I} \psi$}
                \State $\langle \mathcal{M}, \{u^\varphi\} \rangle$ :=  \Call{slidingWindow}{$\mathcal{M}, \varphi, \mathbf{u}$}
            \EndCase
            \Case{$\psi_1 \until{} \psi_2$}
                \State $\langle \mathcal{M}, \{u^\varphi\} \rangle$ :=  \scalebox{0.95}{\Call{unboundedUntil}{$\mathcal{M}, \varphi, \mathbf{u}$}}
            \EndCase
        \EndSwitch
        \State \Return  $\langle \mathcal{M}, \{u^\varphi\} \rangle$
        \EndProcedure
    \end{algorithmic}
\end{algorithm}

\noindent{\bf Online Monitoring Of Non-temporal Operators}
When monitoring formulae containing Boolean or Spatial operators, the online evaluation can be performed very efficiently by simply updating the robustness signal at the times corresponding to the received update. Algorithm~\ref{alg:atoms} shows the algorithm for monitoring atomic formulae, Algorithm~\ref{alg:unary} presents the one for monitoring unary operators (i.e. $\lnot$ and $\escape{d}{}$), and Algorithm~\ref{alg:binary} shows the one for binary operators (i.e. $\lor$ and $\reach{d}{}$).
We represent by \Call{compute\_op}{$\mathbf{OP}, \mathbf{V}$} (and \Call{compute\_op}{$\mathbf{OP}, \mathbf{V}_1, \mathbf{V}_2$}) the execution of the semantic operation corresponding to the operator $\mathbf{OP}$, along the lines of Definitions~\ref{def:boolean-semantics},\ref{def:interval-semantics}, i.e. $[\max]/[\min]$ direct computation for Booleans, and the classical reach/escape routines~\cite{Bartocci17memocode} for spatial operators. A key difference from the offline version of the spatial algorithms, however, is that in our online version the \Call{compute\_op}{} implementation has been crafted to enable spatial-parallelization, i.e. monitors' users with appropriate hardware and the need to speed-up for large spaces, can opt-in for the multi-threaded version of the algorithm, where \Call{compute\_op}{} is executed in parallel for any location of the spatial model.

\begin{algorithm}
    \caption{Atomic Formula Monitoring}
    \begin{algorithmic}[1]
        \Function{atom}{$\mathcal{M}, p \circ c, \mathbf{u}$}
        \State $(t_a, t_b, \mathbf{V}) := \mathbf{u}$
        \State $\mathbf{u}^{p \circ c} :=  (t_a, t_b, \Call{compute\_op}{p \circ c, \mathbf{V})}$
        \State \Call{refine}{$\mathcal{M}[\rho^{p \circ c}], \{\mathbf{u}^{p \circ c}\}$}
        \State \Return $\langle \mathcal{M}, \{\mathbf{u}^{p \circ c}\} \rangle$ 
        \EndFunction
    \end{algorithmic}
    \label{alg:atoms}
\end{algorithm}

\begin{algorithm}
    \caption{Unary Operator Monitoring}
    \begin{algorithmic}[1]
        \Function{unary}{$\mathcal{M}, \mathbf{OP}\psi, \mathbf{u}$}
        \State $\langle \mathcal{M}, \{\mathbf{u}^\psi\} \rangle$ := \Call{monitor}{$\mathcal{M}, \psi, \mathbf{u}$}
        \State $\{\mathbf{u}^{\mathbf{OP}\psi}\}$ := $\emptyset$
        \For{$(t_a, t_b, \mathbf{V}) \in \{\mathbf{u}^\psi\}$}
        \State \scalebox{0.9}{$\{\mathbf{u}^{\mathbf{OP}\psi}\}$ :=  $\{\mathbf{u}^{\mathbf{OP}\psi}\}~\cup (t_a, t_b, \Call{compute\_op}{\mathbf{OP}, \mathbf{V}})$}
        \EndFor
        \State \Call{refine}{$\mathcal{M}[\rho^{\mathbf{OP}\psi}], \{\mathbf{u}^{\mathbf{OP}\psi}\}$}
        \State \Return $\langle \mathcal{M}, \{\mathbf{u}^{\mathbf{OP}\psi}\} \rangle$ 
        \EndFunction
    \end{algorithmic}
    \label{alg:unary}
\end{algorithm}

The algorithm for monitoring binary operators is slightly more complex, as it requires to take into account the corresponding value of the other subformula when an update is processed.
In this context, we indicate by 
\Call{select}{$\mathbf{s}, t_1, t_2$} the restriction of the signal $\mathbf{s}$  to the time interval that starts at $t_1$ and ends at $t_2$ (excluded). 

\begin{algorithm} 
    \caption{Binary Operator Monitoring}
    \begin{algorithmic}[1]
        \Function{binary}{$\mathcal{M}, \psi_1\mathbf{OP}\psi_2, \mathbf{u}$}
        \State $\{\mathbf{u}^{\psi_1\mathbf{OP}\psi_2}\}$ := $\emptyset$
        \State $\langle \mathcal{M}, \{\mathbf{u}^{\psi_1}\} \rangle$ := \Call{monitor}{$\mathcal{M}, \psi_1, \mathbf{u}$}
        
        \For{$(t_a, t_b, \mathbf{V}^{\psi_1}) \in \{\mathbf{u}^{\psi_1}\}$}
            \State $\mathbf{V}^{\psi_2}$ := \Call{select}{$\mathcal{M}[\rho^{\psi_2}], t_a, t_b$} 
            \State $\{\mathbf{u}^{\psi_1\mathbf{OP}\psi_2}\}$ :=  $\{\mathbf{u}^{\psi_1\mathbf{OP}\psi_2}\}$ $\cup$ 
            \\\qquad\qquad\qquad\qquad\quad 
            \scalebox{0.9}{$(t_a, t_b, \Call{compute\_op}{\mathbf{OP}, \mathbf{V}^{\psi_1}, \mathbf{V}^{\psi_2}}$}
        \EndFor
        \State \textit{Repeat lines 3-8 symmetrically for $\psi_2$...}
        \State \Call{refine}{$\mathcal{M}[\rho^{\psi_1\mathbf{OP}\psi_2}], \{\mathbf{u}^{\psi_1\mathbf{OP}\psi_2}\}$}
        \State \Return $\langle \mathcal{M}, \{\mathbf{u}^{\psi_1\mathbf{OP}\psi_2}\} \rangle$ 
        \EndFunction
    \end{algorithmic}
    \label{alg:binary}
\end{algorithm}


\noindent{\bf Online Monitoring Of Temporal Operators}
To execute temporal operators quickly enough for online needs, on the other hand, 
we need to store some extra information throughout the process. 
Firstly, it is useful to recollect that, in general, every temporal operator can be decomposed~\cite{EfficientSTL} in the conjunction of two (efficiently computable) operators:
\begin{itemize}
    \item the bounded eventually $\ev{I}$ (or equivalently the bounded globally $\glob{I}$)
    \item the unbounded until $\until{}$
\end{itemize}
We propose here an enhanced algorithm for monitoring bounded globally/eventually operators with out-of-order updates. For that aim, we slightly adapted the classical  sliding window algorithm from Lemire~\cite{Lemire2006StreamingMF} so that it is constrained on the $\mathfrak{ur}$ function and that it can deal seamlessly with numerical and interval values. Algorithm~\ref{alg:sliding-window} presents 
primary routine of the sliding window for computing updates of bounded unary temporal operators $\mathbf{OP}_I$. 
The algorithm exploits an additional data structure $\mathtt{W}$ that is a deque, such that new elements of the window are added at the end, and such that when the window is saturated (i.e. the elements inside denote a time span bigger than the definition interval $I$ of the operator), they are removed from left and propagated as updates.
The logic of the algorithm is essentially the following: for each update received in input, the sliding window is initialized on the fragment of the robustness signal of the subformula defined by the update ripple function $\mathfrak{ur}$. For each piece of the fragment, the sliding window is updated (line~\ref{lst:line:add}), and each time the new piece makes the data in the window exceed the maximum size, the sliding window slides to the right, removing from the window some elements that can be safely propagated as updates (line~\ref{lst:line:slide}); some edge cases are not covered to keep the algorithm concise (e.g. the case when the current piece is by itself wider than the window size). 
The precise behaviour of \Call{slide}{} and \Call{add}{}, that control the mutation of $\mathtt{W}$ can be examined in the extended version of the paper
or in the Moonlight implementation.

\begin{algorithm}
\caption{Sliding Window}\label{alg:sliding-window}
    \begin{algorithmic}[1]
        \Procedure{slidingWindow}{$\mathcal{M}, \mathbf{OP}_{[t_a, t_b]}\psi, \mathbf{u}$}
        \State $\langle \mathcal{M}, \{\mathbf{u}^\psi\} \rangle$ := \Call{monitor}{$\mathcal{M}, \psi, \mathbf{u}$}
        \State $\{\mathbf{u}^{\mathbf{OP}\psi}\}$ := $\emptyset$
        \For{$\mathbf{u}^\psi \in \{\mathbf{u}^\psi\}$}
            \State $(t_s, t_e) := \mathfrak{ur}(\mathbf{u}^\psi, \mathbf{OP}_{[t_a,t_b]}\psi)$
            \For{$(t, \mathbf{V}) \in \Call{select}{\mathcal{M}[\rho^{\psi}], t_s, t_e} $}
                \If{$t > t_b + \mathtt{W.first.start} $}
                    \State $\{\mathbf{u}^{\mathbf{OP}\psi}\} := \{\mathbf{u}^{\mathbf{OP}\psi}\}  \cup \Call{Slide}{t - t_b}$ \label{lst:line:slide}
                \EndIf
                \State \Call{add}{$t - t_a, \mathbf{V}$} \label{lst:line:add}
            \EndFor
            
        \EndFor
        \State \Call{refine}{$\mathcal{M}[\rho^{\mathbf{OP}\psi}], \{\mathbf{u}^{\mathbf{OP}\psi}\}$}
        \State \Return $\langle \mathcal{M}, \{\mathbf{u}^{\mathbf{OP}\psi}\} \rangle$ 
        \EndProcedure
    \end{algorithmic}
\end{algorithm}

The second fundamental temporal algorithm is the one for computing the unbounded until. Unfortunately, being unbounded, any update might require to recompute, in 
the worst case, the whole robustness/satisfaction signal.
In our implementation, we consider 
the algorithm in~\cite{Deshmukh2017}. 
Note that it requires to keep the minimum value of preceding computations of $\varphi_1$ and the maximum value of preceding computations of the whole formula as secondary data structures.
A last remark about the implementation must be made: while the algorithms have been developed with the goal to enable out-of-order execution, all of them have been implemented also in an in-order variant, so that the execution time penalty from not assuming that updates are at the end, does not affect the users of the tool, when the use case of interest allows to.

%% file: sections/eval.tex
\section{Experimental Evaluation}\label{sec:evaluation}
The interval semantics we presented in Section~\ref{sec:logic} and the online (in-order and out-of-order) monitoring strategies  of Section~\ref{sec:monitoring} have been implemented as extensions to the Moonlight tool. 
To showcase the kind of applications where they can be exploited, and to compare the performances with other state-of-the-art approaches, we propose here three different examples: (i) we present and discuss the results of the properties previously introduced, in the context of air quality monitoring;  (ii) we compare the performances of our approach for the evaluation of a temporal property on the Abstract Fuel Control Simulink model from the Breach~\cite{breach} tool; (iii) we compare the performances of the online approach versus the offline version of Moonlight 
on a simulated sensor network adopting the ZigBee protocol. 
All the computations have been executed on an Intel\textsuperscript{\textregistered} Core\textsuperscript{\texttrademark} i7-5820K CPU @ 3.30GHz, 15M cache, 6 cores (12 threads), with 32GB RAM, running Ubuntu\textsuperscript{\textregistered} 20.04.2 LTS, and Matlab\textsuperscript{\texttrademark} R2021a.


\subsection{Use case: Air pollution monitoring}\label{sec:issues}
\begin{figure*}
     \centering
     \begin{subfigure}[b]{0.4\paperwidth}
         \centering
         \includegraphics[width=0.4\paperwidth]{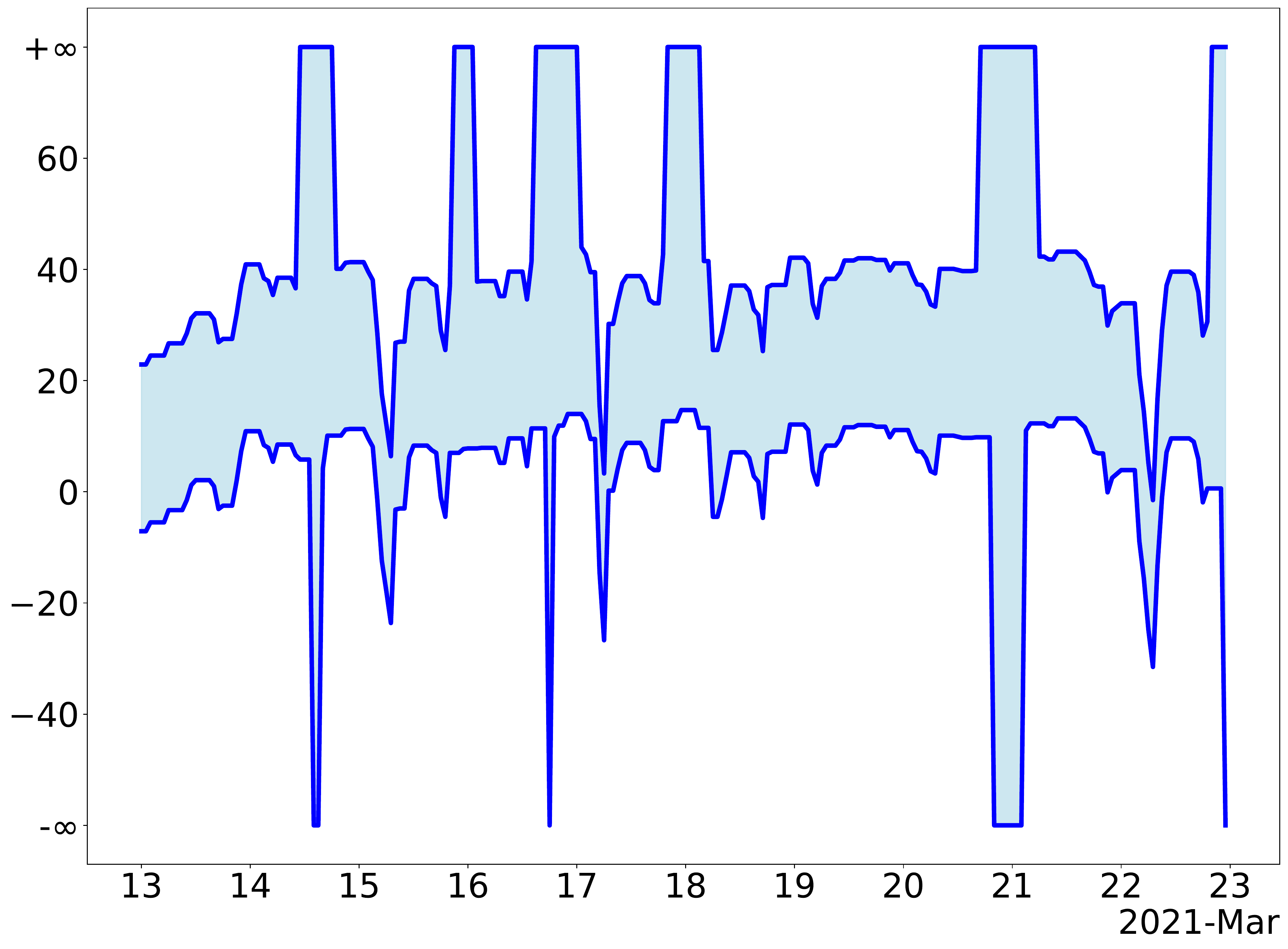}
         \label{fig:rezzato-p1}
     \end{subfigure}
     \hfill
     \begin{subfigure}[b]{0.4\paperwidth}
         \centering
         \includegraphics[width=0.4\paperwidth]{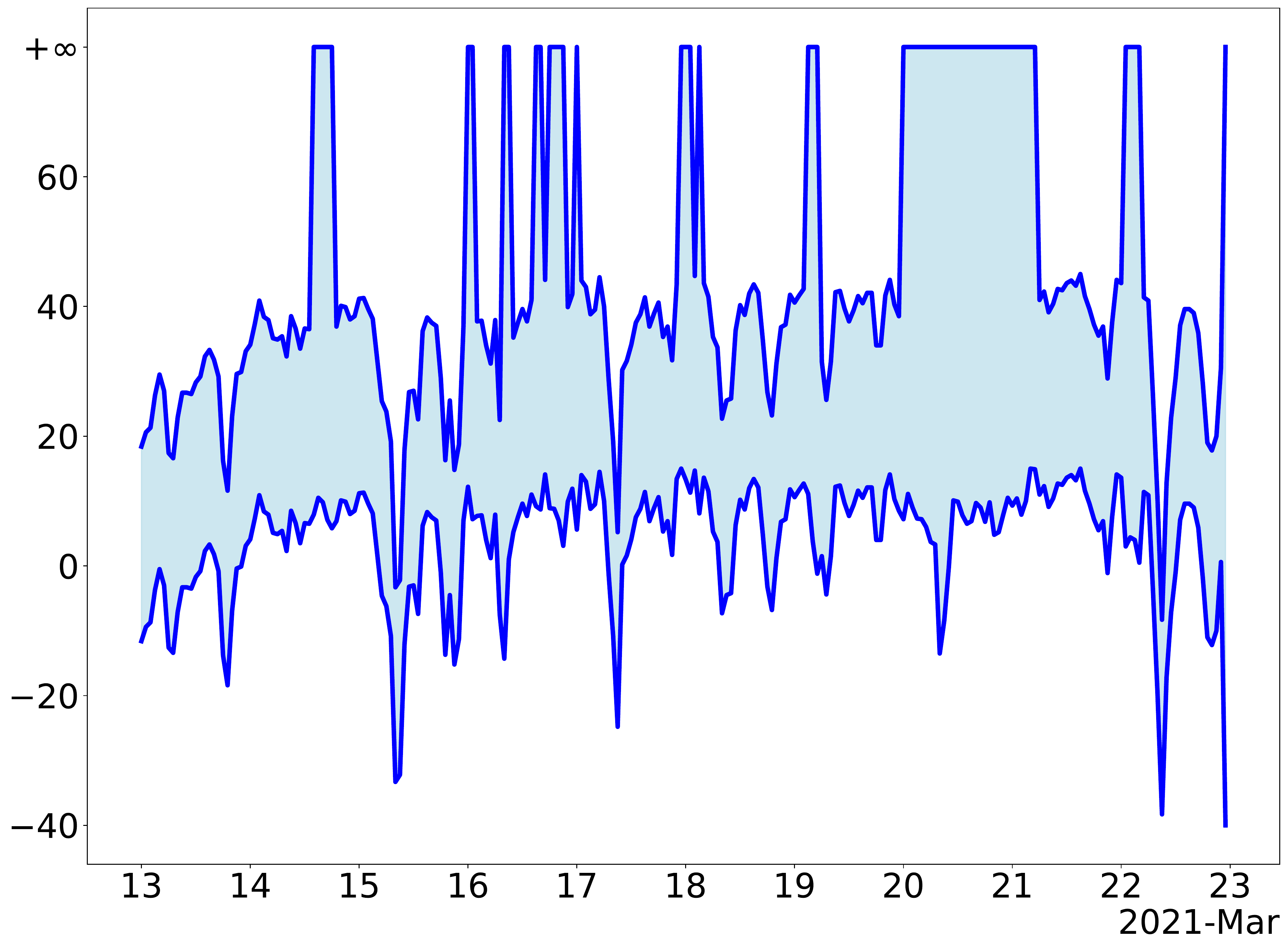}
         \label{fig:rezzato-p2}
     \end{subfigure}
        \caption{
    The results of monitoring robustness for Property~\ref{p1} (on the left), and Property~\ref{p2} (on the right) at the Rezzato station. Despite the missing values, reliable values of the robustness of the property can be provided.}
        \label{fig:rezzato-props}
\end{figure*}

Recalling Properties~\ref{p1}, \ref{p2} from Section~\ref{sec:logic}, we can see in Figure~\ref{fig:rezzato-props} the results of the monitoring. 
Note that when both the upper and lower bounds are below the $0$ threshold, the property is certainly violated, while when only the lower bound is below $0$, then the property is \emph{potentially} violated.
Property~\ref{p1} gives some important insights on the faults observed in Figure~\ref{fig:no2}. In fact, we can see that of the six observed failures for the ten-days span of interest, only three happen for a time that is long enough to potentially trigger public concern, which correspond to the spikes to minus infinity in the lower bound of Figure~\ref{fig:rezzato-props} (left).
In essence, with the interval semantics we learn that the property could potentially be violated in those time-spans, while it is certainly not for the other missing values.
However, Property~\ref{p2} tells us something more about the neighbourhood: in fact, by combining the observations registered from close location, it is apparent that just one of the failures (the one happening during March 20th) likely corresponds to a violation of the property, since there is no close location exhibiting low levels of nitrogen dioxide in Figure~\ref{fig:rezzato-props} (right).

\subsection{Online comparison: Abstract Fuel Control}
Consider a Simulink\textsuperscript{\textregistered} model that describes a black-box representation of an engine's air-fuel ratio controller aimed at complying to emission targets of a vehicle, where the user has direct control over the \textit{engine speed} and \textit{pedal angle}. Each input and output is represented as a signal that is sampled regularly, the outputs being the actual air-to-fuel ratio (AF), and the mean air-to-fuel ratio value for the given input parameters (AF\textsubscript{ref}) at a sampling period $T=0.1s$. For a full description of the model, the reader can see~\cite{afc-original}, while~\cite{Deshmukh2017} provides the reference implementation for online monitoring in Breach.


In our experiments, we monitored the following STL property (note that STREL is an extension of STL, and therefore each STL formula is also a STREL formula)
\begin{equation*}
\small
    \varphi = \glob{[10,30]}(|AF - AF_{ref}| > 0.1 \rightarrow (\ev{[0,1]}|AF-AF_{ref}| < 0.1))
\end{equation*}
for different sample sizes, considering both updates as an order chain and by shuffling them at random to simulate out-of-order retrieval and processing. The result of $|AF-AF_{ref}|$ from the model has been stored in a file and loaded before starting the stopwatch for both monitors, to eliminate the simulation and loading time from the performance evaluation. Breach monitor has been measured via the reference implementation as a Simulink model, while Moonlight is implemented as a Java program. 
Table~\ref{tab:afc} reports a summary of the performances of the monitors for different sample sizes.

\begin{table}[ht]
\centering
\begin{tabular}{|l | l | l | l |}
\hline
N. samples & Breach Exec. Time &  \multicolumn{2}{c|}{Moonlight Exec. Time} \\
 & In-order & In-order& Out-of-order\\
\hline
500 & 7.603 s &  0.004 s & 0.157 s \\
1000 & 8.143 s & 0.016 s & 0.489 s\\
5000 & 10.770 s & 0.096 s & 9.790 s\\
10000 & 13.730 s & 0.113 s & 44.894 s\\
\hline
\end{tabular}
\caption{Performances for monitoring the property $\varphi$ for different sample size. Times are averaged over 100 runs.}
\label{tab:afc}
\end{table}

The interesting insight of the comparison is the fact that, while our in-order implementation provides reliably faster performances (note that the offline version of Moonlight had already shown better performance than Breach in~\cite{BartocciBLNS20}), the penalty that comes from not assuming ordered inputs grows substantially with the increase of the input size, as this requires longer searches in the output signal, to find the spot where the update should be applied.
Nevertheless, the biggest sample size we considered is quite extreme (ten thousand randomly-shuffled samples), yet the execution time (4.489 ms/sample on average) is way smaller than the sampling time (0.1 s), which therefore makes it reasonable for most real-time scenarios. 


\subsection{Moonlight comparison: ZigBee Protocol}
Consider a collection of moving devices communicating via the ZigBee (IEEE 802.15.4) protocol. From the protocol description we know that the devices can have three roles: they can either be coordinators, routers or sensor-node. Each device is equipped with an humidity sensor $h(t)$ that reports at each time instant the observed value of the humidity at the current location. The humidity observed can be described as an MA(0) process, i.e.
\begin{equation*}
    h(t) = c_0 + \varepsilon(t), \tag{$\varepsilon \sim WN(0, \lambda^2)$}
\end{equation*}

where the observed value comes from the real value $c_0$, with some  perturbation from the zero-mean white noise $\varepsilon$ of variance $\lambda^2$. 
Each device can communicate with the ones that are close enough directly, but they can also communicate with furthest ones, as long as there is some router between them that can bridges the communication. 
Let $X_H$ denote the true value of humidity for a given device at a given time, let $X_S$ denote the role of a given sensor, and $T$ some time threshold to warn the observers. We monitored the following properties on the system:
\begin{equation*}
    \varphi_1 = (X_H > 60) \rightarrow \ev{[0, T]} (X_H < 30)
\end{equation*}
\begin{equation*}
    \varphi_2 = \everywhere{}{}\somewhere{d < 10}{}(X_S= \mathtt{coordinator})
\end{equation*}
Property $\varphi_1$ denotes an alert condition: if the humidity measured by a device $X_H$ goes beyond $60\%$, then it must fall down at $30\%$ afterwords, within the time threshold $T$.
Property $\varphi_2$, on the other hand, defines a reachability criterion between the sensors: it checks whether it is true that from any location, it is possible to reach a coordinator ($X_S = \mathtt{coordinator}$) in less than 10 hops. 
Similarly to previous versions of this model~\cite{Bartocci17memocode, tutorial}, we can consider the spatial model as a graph where all the devices are the nodes, and the edges between the nodes are all labeled by $d=1$ to denote the networking hop from one device to another.
Table~\ref{tab:sensors} shows the difference in monitoring $\varphi_1$ and $\varphi_2$ both online and offline. It is interesting to see how the different algorithms behave on same formula and data: in fact, the online temporal algorithms are penalized by the complexity added by the fact that some values must be recomputed. Conversely, the online spatial ones benefit from the hypothesis of spatial synchronization of the locations, resulting in a slightly more efficient computations in the case we explored. Lastly, it can be seen that the benefit of parallelization is particularly evident when the number of nodes is strictly smaller than the number of cores of the CPU (10 in our case), while the benefit practically vanishes (actually resulting in more overhead) as the number of parallel threads grows significantly more than the cores available.  

\begin{table}[ht]
\centering
\begin{tabular}{|l | l | l | l | l | l | l | l |}
\hline

Time & N. & \multicolumn{2}{c|}{Offline} &  \multicolumn{2}{c|}{Online} &  \multicolumn{2}{c|}{Online(Parallel)} \\
samples & nodes & $\varphi_1$ & $\varphi_2$ & $\varphi_1$ & $\varphi_2$ & $\varphi_1$ & $\varphi_2$\\
\hline
\multirow{3}{*}{100} & 10 & 9 & 77  & 116 & 29 & 49 &  58 \\
& 50 & 8 & 1028 & 151 & 430 & 84 & 583 \\
& 100 & 15 & 6919 & 197 & 2993 & 137 & 3017 \\
\multirow{3}{*}{500} & 10 & 8 & 200 & 621 & 45 & 461 & 760 \\
& 50 & 17 &  4058 & 1901 & 1783 & 1549 & 2009 \\
& 100 & 25 & 32561 & 3333 & 15641 & 2889 & 15486 \\
\hline
\end{tabular}
\caption{Execution times registered for monitoring $\varphi_1$ and $\varphi_2$ with different versions of Moonlight. Times in ms, averaged over 100 runs.}
\label{tab:sensors}
\end{table}


%% file: sections/future.tex
\section{Conclusions \& Future Work}\label{sec:future}
We extended the traditional definition of signals to also consider imprecise signals defined by intervals of values. We presented an interval semantics for STREL, we proved its soundness and correctness, and we introduced an online monitoring algorithm for STREL that exploits imprecise signals that can be refined by updates arriving in any order, and that can monitor updates on different locations in parallel. We implemented the proposed methodology in the Moonlight monitoring tool. We motivated our framework from an air pollution control specification with real data from the region of Lombardy, Italy. Lastly, we compared the new methodology with other state-of-the art tools, and discussed the differences.
Many directions of future work can be followed, for example, the \emph{space-synchronization} hypothesis helped us simplifying the implementation of the algorithms, but is not needed from a theoretical point of view. It will be interesting in the future to clearly assess the computational advantages and disadvantages of that hypothesis, and to which extent it can be relaxed. Another intriguing topic for future development concerns spatial models representing (and interacting as) distributed systems. In that context, multiple directions could be pursued, like considering an ownership model for the atomic formulae, or by reasoning on an actor-based communication model among locations. Another interesting idea could be to expand the kind of failures we can monitor, for example, we could consider some form of error correction in case some received updates later prove to have provided wrong information (maybe because of some broken sensors).
Lastly, different form of computational optimization could be explored, like stopping when some bounds on the satisfiability/robustness have been reached, as well as intensive parallelization and hardware acceleration of the main algorithms.

%% file: sections/proofs.tex
\section{Proofs}
\subsection{Proof of Theorem~\ref{th:soundness}}\label{proof:soundness}
\input{proofs/soundness}

\subsection{Proof of Lemma~\ref{th:metric-lemma}}\label{proof:metric-lemma}
\input{proofs/metriclemma}

\subsection{Proof of Theorem~\ref{th:correctness}}\label{proof:correctness}
\input{proofs/correctness}

%% file: proofs/soundness.tex
\begin{proof}
The theorem can be proved by induction on the subformulae of the formula $\varphi$.
\begin{itemize}
    \item $\varphi \equiv \top$: immediate from semantics, since $\rho(\mathbf{s}, \ell, t,\top) > 0$, and $\chi(\mathbf{s},\ell,t,\top) = 1$  for any $\mathbf{s}, \ell, t$. 
    \item $\varphi \equiv \bot$: immediate from semantics, since $\rho(\mathbf{s}, \ell, t,\bot) < 0$, and $\chi(\mathbf{s},\ell,t,\bot) = -1$  for any $\mathbf{s}, \ell, t$. 
    
    \item $\varphi \equiv p \circ c$: immediate from semantics, in fact, we can distinguish two cases based on $\circ$. (i) let $\circ \equiv `> \text{'}$, $\rho(\mathbf{s}, \ell, t, p \circ c) > 0$ iff $\pi_p(\mathbf{s}(\ell,t)) - c > 0$, but this means $\pi_p(\mathbf{s}(\ell,t)) > c$, which implies $\chi(\mathbf{s},\ell,t,p \circ c) = 1$; conversely $\rho(\mathbf{s}, \ell, t, p \circ c) < 0$ iff $\pi_p(\mathbf{s}(\ell,t))  - c < 0$, but this means $\pi_p(\mathbf{s}(\ell,t)) < c$, which implies $\chi(\mathbf{s},\ell,t,p \circ c) = -1$. (ii) let $\circ \equiv `< \text{'}$, $\rho(\mathbf{s}, \ell, t, p \circ c) > 0$ iff $ c - \pi_p(\mathbf{s}(\ell,t)) > 0$, but this means $\pi_p(\mathbf{s}(\ell,t)) < c$, which implies $\chi(\mathbf{s},\ell,t,p \circ c) = 1$; conversely $\rho(\mathbf{s}, \ell, t, p \circ c) < 0$ iff $c - \pi_p(\mathbf{s}(\ell,t)) < 0$, but this means $\pi_p(\mathbf{s}(\ell,t)) > c$, which implies $\chi(\mathbf{s},\ell,t,p \circ c) = -1$. The case for $0 \in \rho(\mathbf{s}, \ell, t, p \circ c)$ trivially holds by semantics since $\chi(\mathbf{s}, \ell, t, p \circ c) = 0$ in all other cases. 
    \item $\varphi \equiv \lnot \varphi_1$:
    Let $\rho(\mathbf{s},\ell,t,\varphi) > 0$, from semantics we have that $\rho(\mathbf{s},\ell,t,\lnot \varphi) = - \rho(\mathbf{s},\ell,t,\varphi_1)$ and therefore $\rho(\mathbf{s},\ell,t,\varphi_1) < 0$ . By inductive hypothesis $\chi(\mathbf{s},\ell,t,\varphi_1) = -1$, but then $\chi(\mathbf{s},\ell,t,\varphi) = -\chi(\mathbf{s},\ell,t,\varphi_1) = 1$.
    Let $\rho(\mathbf{s},\ell,t,\varphi) < 0$, from semantics we have that $\rho(\mathbf{s},\ell,t,\lnot \varphi) = - \rho(\mathbf{s},\ell,t,\varphi_1)$ and therefore $\rho(\mathbf{s},\ell,t,\varphi_1) > 0$ . By inductive hypothesis $\chi(\mathbf{s},\ell,t,\varphi_1) = 1$, but then $\chi(\mathbf{s},\ell,t,\varphi) = -\chi(\mathbf{s},\ell,t,\varphi_1) = -1$.
    The case for $0 \in \rho(\mathbf{s},\ell,t,\varphi)$ is similar.
    
    \item $\varphi \equiv \psi_1 \lor \psi_2$: consider \Circled{1} $\equiv [\max](\rho(\mathbf{s}, \ell, t, \psi_1), \rho(\mathbf{s}, \ell, t, \psi_2))$.
    \begin{itemize}
        \item Let $\Circled{1} > 0$:
              by definition, we either have $\rho(\mathbf{s},\ell,t,\psi_1) > 0$ or $\rho(\mathbf{s},\ell,t,\psi_2) > 0$. Then, by inductive hypothesis, the theorem holds on both $\psi_1$ and $\psi_2$, and therefore either $\chi(\mathbf{s},\ell,t,\psi_1) = 1$ or $\chi(\mathbf{s},\ell,t,\psi_2) = 1$. But, by semantics we have that $\chi(\mathbf{s},\ell,t,\varphi) = \max(\chi(\mathbf{s},\ell,t,\psi_1),$   $\chi(\mathbf{s},\ell,t,\psi_2)) = 1$, and therefore the theorem holds.
        \item Let \Circled{1} $< 0$:
              by definition, we must have both $\rho(\mathbf{s},\ell,t,\psi_1) < 0$ and $\rho(\mathbf{s},\ell,t,\psi_2) < 0$. Then, by inductive hypothesis, the theorem holds on both $\psi_1$ and $\psi_2$, and therefore $\chi(\mathbf{s},\ell,t,\psi_1) = \chi(\mathbf{s},\ell,t,\psi_2) = -1$. But, by semantics we have that $\chi(\mathbf{s},\ell,t,\varphi) = \max(\chi(\mathbf{s},\ell,t,\psi_1),$   $\chi(\mathbf{s},\ell,t,\psi_2)) = -1$, and therefore the theorem holds.
        
    \end{itemize}
    The case for $0 \in \rho(\mathbf{s},\ell,t,\varphi)$ is similar.
    
    \item $\varphi \equiv \psi_1 \until{I} \psi_2$: consider \Circled{2} $\equiv [\max\limits_{t'\in t + I}] \{[\min]($\Circled{3}, \Circled{4}$)\}$, with \Circled{3} $\equiv \rho(\mathbf{s},\ell, t', \psi_2)$ and \Circled{4} $\equiv [\min\limits_{t''\in [t', t]}]\{\rho(\mathbf{s},\ell, t'', \psi_1)\}$.
    \begin{itemize}
        \item Let \Circled{2} $> 0$: there will be some $t' \in t + I$ such that $[\min]($\Circled{3}, \Circled{4}$) > 0$.This means that both \Circled{3} $> 0$ and \Circled{4} $> 0$.
              However, by inductive hypothesis, this means that $\chi(\mathbf{s},\ell,t',\psi_2) = \chi(\mathbf{s},\ell,t'',\psi_1) = 1$, for all $t'' \in [t', t]$. Therefore, by semantics we have that $\chi(\mathbf{s},\ell,t',\varphi) = 1$, and the theorem holds.
        \item Let \Circled{2} $< 0$: we can partition the set $t+I$ in two cases:
        \begin{itemize}
            \item $t'\in t+I$ s.t. \Circled{2} $ = $ \Circled{3}; 
                  by inductive hypothesis the theorem holds for $\psi_2$, therefore $\chi(\mathbf{s},\ell,t',\psi_2) = -1$. In this case, by semantics, we have that $\chi(\mathbf{s},\ell,t,\varphi) = -1$, regardless of the value of \Circled{4}.
                  Therefore the theorem holds.
            \item $t'\in t+I$ s.t. \Circled{2} $ = $ \Circled{4};
                  there is some $t'' \in [t', t]$ s.t. $\rho(\mathbf{s}, t'', \psi_1) < 0$. By inductive hypothesis the theorem holds for $\psi_1$, therefore $\chi(\mathbf{s},\ell,t'',\psi_1) = -1$. In this case, by semantics, we have that $\chi(\mathbf{s},\ell,t,\varphi) = -1$ regardless of the value of \Circled{3}. Therefore the theorem holds.
        \end{itemize}
    \end{itemize}
    The case for $0 \in \rho(\mathbf{s},\ell,t,\varphi)$ is similar.
    
    \item $\varphi \equiv \psi_1 \reach{\leq d}{} \psi_2$: the proof is the same as for $\varphi \equiv \psi_1 \until{I} \psi_2$, where the condition $t' \in t + I$ is replaced with $\tau \in Routes(\ell), \ell' \in \tau: d_\tau[\ell'] \in [d_1, d_2]$ and $t'' \in [t', t]$ with $\ell''< \tau(\ell')$.
    \item $\varphi \equiv \escape{\geq d}{} \psi$: from semantics we have that $\Circled{1} \equiv \rho(\mathbf{s}, \ell, t, \escape{\geq d}{} \psi) = \left[\max\limits_{\tau \in Routes(\ell)}\right] \left[\max\limits_{\ell' \in \tau : d_\mathcal{S}[\ell, \ell'] \in [d, \infty]}\right]\{\Circled{2}\}$, with   $\Circled{2} \equiv \left[\min\limits_{\ell'' < \tau(\ell')}\right]\{\rho(\mathbf{s}, \ell'', t, \psi)\}\}$. 
    \begin{itemize}
        \item Let $\Circled{1} > 0$: then there is a $\tau \in Routes(\ell)$, and $\ell' \in \tau$, such that $\Circled{2} > 0$. But this implies that exists an $\ell''<\tau(\ell')$ such that $\rho(s, \ell'',t,\varphi) > 0$. But by inductive hypothesis $\rho{s, \ell'' , t, \varphi} > 0$ implies that $\chi{s, \ell'' , t, \varphi} = 1$, hence the thesis.
        \item Let $\Circled{1} < 0$: then for any $\tau \in Routes(\ell)$, and $\ell' \in \tau$, it must be $\Circled{2} < 0$. But this implies that exists an $\ell''<\tau(\ell')$ such that $\rho(s, \ell'',t,\varphi) < 0$. But by inductive hypothesis $\rho(s, \ell'' , t, \varphi) < 0$ implies that $\chi(s, \ell'' , t, \varphi) = - 1$, hence the thesis.
        \item Let $\Circled{1} \ni 0$: then it must be that there is no $\tau \in Routes(\ell)$ having $\ell' \in \tau$ such that $\Circled{2} > 0$, and that exists a route $\tau$ having a location $\ell'$ such that $\Circled{2} \not< 0$. For this location, we have that $\underline{\Circled{2}} \leq 0$ and that $\overline{\Circled{2}} \geq 0$, which mean that $0 \in  \rho(\mathbf{s}, \ell, t, \varphi)$. So, by inductive hypothesis, we have  $\chi(\mathbf{s}, \ell, t, \varphi) = 0$, hence the thesis.
    \end{itemize}
\end{itemize}
\end{proof}

%% file: proofs/metriclemma.tex
\begin{proof}
The theorem can be proved by induction on the subformulae of the formula $\varphi$.
\begin{itemize}
    \item $\varphi \equiv \top | \bot$: $||\rho_{s_1}^{\varphi}-\rho_{s_2}^{\varphi}||_\infty = 0$ for any $\mathbf{s_1}, \mathbf{s_2}$.
    \item  $\varphi \equiv p~\circ~c$: $||\rho_{s_1}^{ \varphi}-\rho_{s_2}^{\varphi}||_\infty < \delta $ implies that $\max\limits_{i \leq n}\max\limits_{l \in \mathbb{L}}\max\limits_{t \in \mathbb{T}}\\\{d_H(\pi_p(\mathbf{s_1}(\ell, t)), \pi_p(\mathbf{s_2}(\ell, t)))\} < \delta$ for any $p \in AP$, but for hypothesis we know that $\max\limits_{t \in \mathbb{T}}\{d_H(\pi_i(\mathbf{s_1}(\ell, t)), \pi_i(\mathbf{s_2}(\ell, t)))\} < \delta$ for any $i \leq n$.
    \item $\varphi \equiv \lnot \psi$: $||\rho_{s_1}^{ \varphi}-\rho_{s_2}^{\varphi}||_\infty = \max\limits_{l \in \mathbb{L}}\max\limits_{t \in \mathbb{T}}\{d_H(-\rho^{\psi}_{s_1}(\ell, t), -\rho^{\psi}_{s_2}(\ell, t)\}$ $=\max\limits_{l \in \mathbb{L}}\max\limits_{t \in \mathbb{T}}\{d_H(\rho^{\psi}_{s_1}(\ell, t), \rho^{\psi}_{s_2}(\ell, t))\}= ||\rho_{s_1}^{\psi}-\rho_{s_2}^{ \psi}||_\infty$. For inductive hypothesis we have that $||\rho_{s_1}^{\psi}-\rho_{s_2}^{\psi}||_\infty < \delta$ for $||s_1-s_2||_\infty < \delta$, therefore, the theorem holds. 
    \item $\varphi \equiv \psi_1 \lor \psi_2$:
    $||\rho_{s_1}^{\varphi}-\rho_{s_2}^{\varphi}||_\infty = \max\limits_{l \in \mathbb{L}}\max\limits_{t \in \mathbb{T}}\{d_H(\Circled{1}, \Circled{2})\} < \delta$ if for any $\ell \in \mathbb{L}$, $t \in \mathbb{T}$, $\max(|\underline{\Circled{1}}-\underline{\Circled{2}}|, |\overline{\Circled{1}}-\overline{\Circled{2}}|) < \delta$, with \Circled{1} $\equiv [\max](\rho_{s_1}^{\psi_1}(\ell, t), \rho_{s_1}^{\psi_2}(\ell, t))$, and  \Circled{2} $\equiv [\max](\rho_{s_2}^{\psi_1}(\ell, t), \rho_{s_2}^{\psi_2}(\ell, t))$.
    The proof can be split in two cases:
    \begin{enumerate}
        \item $|\underline{\Circled{1}}-\underline{\Circled{2}}| < \delta$:
        at any time instant $t$, we will have four cases, depending on the values between $\underline{\rho_{s_1}^{\psi_1}}(\ell, t)$, $\underline{\rho_{s_1}^{\psi_2}}(\ell, t)$, $\underline{\rho_{s_2}^{\psi_1}}(\ell, t)$ and $\underline{\rho_{s_2}^{\psi_2}}(\ell, t)$, which dominate the $[\max]$ operator.
        \begin{enumerate}
            \item $|\underline{\rho_{s_1}^{\psi_1}}(\ell, t) - \underline{\rho_{s_2}^{\psi_1}}(\ell, t)| < \delta$ with $\underline{\rho_{s_1}^{\psi_1}}(\ell, t) \geq \underline{\rho_{s_1}^{\psi_2}}(\ell, t)$ and $\underline{\rho_{s_2}^{\psi_1}}(\ell, t) \geq \underline{\rho_{s_2}^{\psi_2}}(\ell, t)$. Then for inductive hypothesis the theorem holds.
            \item $|\underline{\rho_{s_1}^{\psi_2}}(\ell, t) - \underline{\rho_{s_2}^{\psi_2}}(\ell, t)| < \delta$ with $\underline{\rho_{s_1}^{\psi_2}}(\ell, t) \geq \underline{\rho_{s_1}^{\psi_1}}(\ell, t)$ and $\underline{\rho_{s_2}^{\psi_2}}(\ell, t) \geq \underline{\rho_{s_2}^{\psi_1}}(\ell, t)$. Then for inductive hypothesis the theorem holds.
            
            \item $|\underline{\rho_{s_1}^{\psi_1}}(\ell, t) - \underline{\rho_{s_2}^{\psi_2}}(\ell, t)| < \delta$ with $\underline{\rho_{s_1}^{\psi_1}}(\ell, t) \geq \underline{\rho_{s_1}^{\psi_2}}(\ell, t)$ and $\underline{\rho_{s_2}^{\psi_2}}(\ell, t) \geq \underline{\rho_{s_2}^{\psi_1}}(\ell, t)$. 
            Either (i)~$\underline{\rho_{s_1}^{\psi_1}}(\ell, t) < \underline{\rho_{s_2}^{\psi_2}}(\ell, t) + \delta$, which, combined with case's hypothesis, implies that $\underline{\rho_{s_2}^{\psi_2}}(\ell, t) > \underline{\rho_{s_1}^{\psi_1}}(\ell, t) - \delta \geq \underline{\rho_{s_1}^{\psi_2}}(\ell, t) - \delta$, and therefore $\underline{\rho_{s_2}^{\psi_2}}(\ell, t) - \underline{\rho_{s_1}^{\psi_2}}(\ell, t) > \delta$, which is true by inductive hypothesis.
            Or (ii)~$\underline{\rho_{s_1}^{\psi_1}}(\ell, t) > \underline{\rho_{s_2}^{\psi_2}}(\ell, t) - \delta$,
            which, combined with case's hypothesis, implies that $\underline{\rho_{s_1}^{\psi_1}}(\ell, t) > \underline{\rho_{s_2}^{\psi_2}}(\ell, t) - \delta \geq \underline{\rho_{s_2}^{\psi_1}}(\ell, t) - \delta$, and therefore $\underline{\rho_{s_1}^{\psi_1}}(\ell, t) - \underline{\rho_{s_2}^{\psi_1}}(\ell, t) > - \delta$, which is true by inductive hypothesis.
            
            \item $|\underline{\rho_{s_1}^{\psi_2}}(\ell, t) - \underline{\rho_{s_2}^{\psi_1}}(\ell, t)| < \delta$ with $\underline{\rho_{s_1}^{\psi_2}}(\ell, t) \geq \underline{\rho_{s_1}^{\psi_1}}(\ell, t)$ and $\underline{\rho_{s_2}^{\psi_1}}(\ell, t) \geq \underline{\rho_{s_2}^{\psi_2}}(\ell, t)$. This case is analogous to the previous one, with the robustness signals switched.
        \end{enumerate}
     
        \item $|\overline{\Circled{1}}-\overline{\Circled{2}}| < \delta$. This case is the dual of the previous one, where $\underline{~\cdot~}$ is substituted with $\overline{~\cdot~}$.
        
    \end{enumerate} 
    
    \item $\varphi \equiv \psi_1 \until{I} \psi_2$:
    $||\rho_{s_1}^{\varphi}-\rho_{s_2}^{\varphi}||_\infty =  \max\limits_{\ell \in \mathbb{L}}\max\limits_{t \in \mathbb{T}}\{d_H(\Circled{1}, \Circled{2})\} < \delta$, which requires that for any $\ell \in \mathbb{L}, t \in \mathbb{T}$, both  $|\underline{\Circled{1}}-\underline{\Circled{2}}| < \delta$, and $ |\overline{\Circled{1}}-\overline{\Circled{2}}| < \delta$, with \Circled{1} $\equiv [\max\limits_{t'\in I + t}][\min]\left(\rho^{\psi_2}_{s_1}(\ell, t'), [\min\limits_{t''\in[t,t']}]\{\rho^{\psi_1}_{s_1}(\ell, t'')\}\right)$, and  \Circled{2} $\equiv [\max\limits_{t'\in I + t}][\min]\left(\rho^{\psi_2}_{s_2}(\ell, t'), [\min\limits_{t''\in[t,t']}]\{\rho^{\psi_1}_{s_2}(\ell, t'')\}\right)$.
    
    \begin{enumerate}
        \item  $|\underline{\Circled{1}}-\underline{\Circled{2}}| < \delta$:
           for any $\ell \in \mathbb{L}$, for any $t \in \mathbb{T}$, we have that $|\underline{\Circled{1}}-\underline{\Circled{2}}| \leq \max\limits_{t'\in I + t} \underline{\Circled{A}} < \delta$ iff for all $t' \in I + t$, \underline{\Circled{A}} $< \delta$, with \Circled{A} $\equiv |[\min] (\rho^{\psi_2}_{s_1}(\ell, t'), \left[\min\limits_{t'' \in [t, t']}\right]\{\rho^{\psi_1}_{s_1}(\ell, t'')\}) - \min (\rho^{\psi_2}_{s_2}(\ell, t'), \min\limits_{t'' \in [t, t']}\{\rho^{\psi_1}_{s_2}(\ell, t'')\})|$ .
           This can happen in four subcases:
           \begin{enumerate}
               \item \underline{\Circled{A}} $= | \underline{\rho^{\psi_2}_{s_1}}(\ell, t') -  \underline{\rho^{\psi_2}_{s_2}}(\ell, t'))|$, in which case, it must also be that $ \underline{\rho^{\psi_2}_{s_1}}(\ell, t') \leq \min\limits_{t'' \in [t, t']} \underline{\rho^{\psi_1}_{s_1}}(\ell, t'')$, and $ \underline{\rho^{\psi_2}_{s_2}}(\ell, t') \leq \min\limits_{t'' \in [t, t']} \underline{\rho^{\psi_1}_{s_2}}(\ell, t'')$. But in this case $ \underline{\Circled{A}} < \delta$ by inductive hypothesis. 
               \item  \underline{\Circled{A}} $= |\min\limits_{t'' \in [t, t']} \underline{\rho^{\psi_1}_{s_1}}(\ell, t'') -  \min\limits_{t'' \in [t, t']} \underline{\rho^{\psi_1}_{s_2}}(\ell, t'')|$, in which case, it must also be that $\underline{\rho^{\psi_2}_{s_1}}(\ell, t') \geq \min\limits_{t'' \in [t, t']}\underline{\rho^{\psi_1}_{s_1}}(\ell, t'')$, and $\underline{\rho^{\psi_2}_{s_2}}(\ell, t') \geq \min\limits_{t'' \in [t, t']}\underline{\rho^{\psi_1}_{s_2}}(\ell, t'')$. But then again $\Circled{A} < \delta$ by combination of inductive hypothesis and minimum property.
               \item  \underline{\Circled{A}} $= | \underline{\rho^{\psi_2}_{s_1}}(\ell, t') -  \min\limits_{t'' \in [t, t']} \underline{\rho^{\psi_1}_{s_2}}(\ell, t'')|$, with $\underline{\rho^{\psi_2}_{s_1}}(\ell, t') \leq \min\limits_{t'' \in [t, t']}\underline{\rho^{\psi_1}_{s_1}}(\ell, t'')$, and $\underline{\rho^{\psi_2}_{s_2}}(\ell, t') \leq \min\limits_{t'' \in [t, t']}\underline{\rho^{\psi_1}_{s_2}}(\ell, t'')$  . This case happens iff two conditions hold: 
                  \begin{enumerate}
                      \item $\min\limits_{t'' \in [t, t']} \underline{\rho^{\psi_1}_{s_2}}(\ell, t'') <  \underline{\rho^{\psi_2}_{s_1}}(\ell, t') + \delta$. But in this case, we also have that $\min\limits_{t'' \in [t, t']} \underline{\rho^{\psi_1}_{s_2}}(\ell, t'') < \underline{\rho^{\psi_2}_{s_1}}(\ell, t') < \underline{\rho^{\psi_2}_{s_1}}(\ell, t') + \delta$, where the first inequality holds  by hypothesis of current subcase, while the latter for inductive hypothesis.
                      \item $\min\limits_{t'' \in [t, t']} \underline{\rho^{\psi_1}_{s_2}}(\ell, t'') > \underline{\rho^{\psi_2}_{s_1}}(\ell, t') - \delta$, but then we also have that also $ \underline{\rho^{\psi_2}_{s_1}}(\ell, t') - \delta \leq \min\limits_{t'' \in [t, t']} \underline{\rho^{\psi_1}_{s_1}}(\ell, t'')$ $< \min\limits_{t'' \in [t, t']} \underline{\rho^{\psi_1}_{s_2}}(\ell, t'')  - \delta + \delta$, where the first inequality holds by hypothesis of current subcase, while the latter for inductive hypothesis combined with the property of the $\min$ operator.
                  \end{enumerate}
               \item  \underline{\Circled{A}} $= |\min\limits_{t'' \in [t, t']} \underline{\rho^{\psi_1}_{s_1}}(\ell, t'') -  \underline{\rho^{\psi_2}_{s_2}}(\ell, t')|$: symmetrical to previous case.
           \end{enumerate}
            
        \item  $|\overline{\Circled{1}}-\overline{\Circled{2}}| < \delta$: symmetrical to previous case.
    \end{enumerate}
    

    \item $\varphi \equiv \psi_1 \reach{\leq d}{} \psi_2$: the proof follows symmetrically from the same scheme as for the previous case, except that the role of $\ell$ and $t$ are inverted. 
    
    \item $\varphi \equiv \escape{\geq d}{} \psi$: $||\rho_{s_1}^{ \varphi}-\rho_{s_2}^{\varphi}||_\infty = \max\limits_{l \in \mathbb{L}}\max\limits_{t \in \mathbb{T}}\{d_H(\Circled{1}, \Circled{2}\}$, with $\Circled{1} \equiv  \left[\max\limits_{\tau \in Routes(\ell)}\right] \left[\max\limits_{\ell' \in \tau : d_\mathcal{S}[\ell, \ell'] \in [d, \infty]}\right]\rho_{s_1}^{ \psi}$ and $\Circled{2} \equiv  \left[\max\limits_{\tau \in Routes(\ell)}\right] \left[\max\limits_{\ell' \in \tau : d_\mathcal{S}[\ell, \ell'] \in [d, \infty]}\right]\rho_{s_2}^{ \psi}$. 
    From induction hypothesis, we know that for any $\ell \in \mathbb{L}, t \in \mathbb{T}$, we have that $||\rho_{s_1}^{ \varphi}-\rho_{s_2}^{\varphi}||_\infty < \delta$, so it must necessarily also be that $||\rho_{s_1}^{ \varphi}-\rho_{s_2}^{\varphi}||_\infty < \delta$.
\end{itemize}
\end{proof}

%% file: proofs/correctness.tex
\begin{proof}
The proof can be split in three possible cases for the hypothesis, depending on the value of $|\rho(\mathbf{s_1}, \ell, t, \varphi)|$. 
\begin{itemize}
    \item Let $\rho(\mathbf{s_1}, \ell, t, \varphi) > 0$: we must then have that, $||\mathbf{s_1}-\mathbf{s_2}||_\infty$  $< \underline{\rho}(\mathbf{s_1}, \ell, t, \varphi)$. 
    Moreover, by Theorem~\ref{th:soundness} we have that $\chi(\mathbf{s_1},t,\varphi) = 1$.
    By Lemma~\ref{th:metric-lemma} we have that for $\delta = \underline{\rho}(\mathbf{s_1}, \ell, t, \varphi)$:
    $||\rho_{s_1}^{\varphi}(t) - \rho_{s_2}^{\varphi}(t)||_\infty < \underline{\rho}(\mathbf{s_1}, \ell, t, \varphi)$, but this implies that $d_H(\rho(\mathbf{s_1}, \ell, t, \varphi), \rho(\mathbf{s_2}, \ell, t, \varphi)) < \underline{\rho}(\mathbf{s_1}, \ell, t, \varphi)$, from the Definition~\ref{def:signal-distance}. From definition of $d_H$, this also implies that
    $ |\underline{\rho}(\mathbf{s_1}, \ell, t, \varphi) -  \underline{\rho}(\mathbf{s_2}, \ell, t, \varphi)|< \underline{\rho}(\mathbf{s_1}, \ell, t, \varphi)$. 
    But this also implies that  $\underline{\rho}(\mathbf{s_1,} t, \varphi) -  \underline{\rho}(\mathbf{s_2}, \ell, t, \varphi) < \underline{\rho}(\mathbf{s_1}, \ell, t, \varphi)$, and therefore that  $ \underline{\rho}(\mathbf{s_2}, \ell, t, \varphi) >0$, which by Theorem~\ref{th:soundness} implies $\chi(\mathbf{s_2},t,\varphi) = 1$.
    
    \item Let $\rho(\mathbf{s}_1, \ell, t, \varphi) < 0$: this case is symmetric to the first one, for $\delta = -\overline{\rho}(\mathbf{s}_1, \ell, t, \varphi)$.
    \item Let $0 \in \rho(\mathbf{s}_1, \ell, t, \varphi)$: we have that $||\mathbf{s_1}-\mathbf{s_2}||_\infty < \min(\overline{\rho}(\mathbf{s_1}, \ell, t, \varphi),$ $ - \underline{\rho}(\mathbf{s_1}, \ell, t, \varphi))$ and that $\rho(\mathbf{s_1}, \ell, t, \varphi)) = 0$. 
    \begin{itemize}
        \item[$*$] Choosing $\delta := \overline{\rho}(\mathbf{s_1}, \ell, t, \varphi)$, from the metric lemma we have that  $||\rho_{s_1}^{\varphi}(t) - \rho_{s_2}^{\varphi}(t)||_\infty < \overline{\rho}(\mathbf{s_1}, \ell, t, \varphi)$, but this implies that $d_H(\rho(\mathbf{s_1}, \ell, t, \varphi), \rho(\mathbf{s_2}, \ell, t, \varphi)) < \overline{\rho}(\mathbf{s_1}, \ell, t, \varphi)$, from the Definition~\ref{def:signal-distance}. From definition of $d_H$, this also implies that $ |\overline{\rho}(\mathbf{s_1}, \ell, t, \varphi) -  \overline{\rho}(\mathbf{s_2}, \ell, t, \varphi)|< \overline{\rho}(\mathbf{s_1}, \ell, t, \varphi)$.  
        But this also implies that  $\overline{\rho}(\mathbf{s_1}, \ell t, \varphi) <  \overline{\rho}(\mathbf{s_1}, \ell, t, \varphi) + \overline{\rho}(\mathbf{s_2}, \ell, t, \varphi)$, and therefore that  $\overline{\rho}(\mathbf{s_2}, \ell, t, \varphi) > 0$. 
        
        \item[$*$] Choosing $\delta := -\underline{\rho}(\mathbf{s_1}, \ell, t, \varphi)$, from the metric lemma we have that $||\rho_{s_1}^{\varphi}(t) - \rho_{s_2}^{\varphi}(t)||_\infty < -\underline{\rho}(\mathbf{s_1}, \ell, t, \varphi)$, but this implies that $d_H(\rho(\mathbf{s_1}, \ell, t, \varphi), \rho(\mathbf{s_2}, \ell, t, \varphi)) < -\underline{\rho}(\mathbf{s_1}, \ell, t, \varphi)$, from the Definition~\ref{def:signal-distance}. From definition of $d_H$, this also implies that $ |\underline{\rho}(\mathbf{s_1}, \ell, t, \varphi) -  \underline{\rho}(\mathbf{s_2}, \ell, t, \varphi)|< -\underline{\rho}(\mathbf{s_1}, \ell, t, \varphi)$.  
        But this also implies that  $\underline{\rho}(\mathbf{s_1}, \ell t, \varphi) -  \underline{\rho}(\mathbf{s_2}, \ell, t, \varphi) < -\underline{\rho}(\mathbf{s_1}, \ell, t, \varphi)$, and therefore that  $ \underline{\rho}(\mathbf{s_2}, \ell, t, \varphi) < 0$.
    \end{itemize}
    By Theorem~\ref{th:soundness}, since we have $0 \in \rho(\mathbf{s_2}, \ell, t, \varphi)$ then we must have $\chi(\mathbf{s_2},\ell, t,\varphi) = 0$.

\end{itemize}
\end{proof}

%% file: sections/extra_algo.tex
\balance

\section{Algorithms}
Sliding window mutation primitives are described here.

\begin{algorithm}
\caption{Slide}\label{alg:slide}
    \begin{algorithmic}[1]
        \Function{slide}{$t$}
            \State $(t_i, \mathbf{V}_i) := \mathtt{W.removeFirst()}$
            \While{$\mathtt{W.first.start} < t$}
                \State $(t_{i+1}, \mathbf{V}_{i+1}) := \mathtt{W.removeFirst()}$
                \State $\{\mathbf{u}^{\mathbf{OP}\psi}\}$ := $\{\mathbf{u}^{\mathbf{OP}\psi}\} \cup 
                {(t_i, t_{i+1}, \mathbf{V}_i)}$
                \If{$t_{i+1} > t$}
                    \State $\mathtt{W.addFirst}(t, \mathbf{V}_i)$
                    \State $\mathtt{W.addFirst}(t_{i+1}, \mathbf{V}_{i+1})$
                \Else
                    \State $(t_i, \mathbf{V}_i) := (t_{i+1}, \mathbf{V}_{i+1})$
                \EndIf
            \EndWhile
            \State $\{\mathbf{u}^{\mathbf{OP}\psi}\}$ := $\{\mathbf{u}^{\mathbf{OP}\psi}\} \cup 
                {(t_i, t_{i+1}, \mathbf{V}_i)}$
            \State $\mathtt{W.addFirst}(t, \mathbf{V}_i)$
            \State \Return $\{\mathbf{u}^{\mathbf{OP}\psi}\}$
        \EndFunction
    \end{algorithmic}
\end{algorithm}

\begin{algorithm}
\caption{Add}\label{alg:do-add}
    \begin{algorithmic}[1]
        \Procedure{add}{$\mathbf{OP}, t, \mathbf{V}$}
            \State $(t_2,\mathbf{V}_2) := \mathtt{W.removeLast()}$
            \State $\mathbf{V}_3 := $ \Call{compute\_op}{$\mathbf{OP}, \mathbf{V}, \mathbf{V}_2$}
            \If{$\mathbf{V}_2 = \mathbf{V}$}
                \State $\mathtt{W.addLast}(t_2, \mathbf{V}_2)$
            \ElsIf{$\mathbf{V}_3 = \mathbf{V}$}
                \State \Call{add}{$\mathbf{OP}, t_2, \mathbf{V}_3$}
            \ElsIf{$\mathbf{V}_3 \neq \mathbf{V}_2$}
                \State \Call{add}{$\mathbf{OP}, t_2, \mathbf{V}_3$}
                \State $\mathtt{W.addLast}(t, \mathbf{V})$
            \Else 
                \State $\mathtt{W.addLast}(t_2, \mathbf{V}_2)$
                \State $\mathtt{W.addLast}(t, \mathbf{V})$
            \EndIf 
        \EndProcedure
    \end{algorithmic}
\end{algorithm}